\documentclass[amsmath,trackchanges,twocolumn]{aastex702}
\usepackage{newtxtext,newtxmath}
\usepackage[T1]{fontenc}
\usepackage{fix-cm}
\usepackage{xcolor}
\usepackage{placeins}
\usepackage{graphicx}	
\usepackage{amsmath}	
\usepackage{float}
\usepackage{booktabs}
\usepackage{longtable}
\usepackage{hyperref}
\usepackage{subcaption}
\usepackage{afterpage}

\begin{document}

\title{Asteroseismology of the High-amplitude $\delta$ Scuti Star EH~Librae}

\author[orcid=0009-0000-4210-1270,gname=Xiran, sname='Xie']{Xiran Xie}
\affiliation{Institute for Frontiers in Astronomy and Astrophysics, Beijing Normal University, Beijing 102206, People's Republic of China}
\affiliation{School of Physics and Astronomy, Beijing Normal University, Beijing 100875, People's Republic of China}
\email{xxie@mail.bnu.edu.cn}  

\author[orcid=0000-0001-8241-1740,gname=Jianning, sname='Fu']{Jianning Fu} 
\affiliation{Institute for Frontiers in Astronomy and Astrophysics, Beijing Normal University, Beijing 102206, People's Republic of China}
\affiliation{School of Physics and Astronomy, Beijing Normal University, Beijing 100875, People's Republic of China}
\affiliation{Xinjiang Astronomical Observatory, Chinese Academy of Sciences, Urumqi 830011, Xinjiang, People's Republic of China}
\email[show]{jnfu@bnu.edu.cn}
\correspondingauthor{Jianning Fu}

\author[gname=Gang,sname='Meng']{Gang Meng}
\affiliation{Institute for Frontiers in Astronomy and Astrophysics, Beijing Normal University, Beijing 102206, People's Republic of China}
\affiliation{School of Physics and Astronomy, Beijing Normal University, Beijing 100875, People's Republic of China}
\email{981214502@qq.com}

\author[orcid=0000-0002-2781-873X, gname=Lester, sname='Fox Machado']{Lester Fox Machado}
\affiliation{Observatorio Astron\'omico Nacional, Instituto de Astronomía, Universidad Nacional Autónoma de México, A.P. 106, 22800 Ensenada, México}
\email{lfox@astro.unam.mx}

\author[orcid=0000-0003-1263-808X,gname=Ra\'ul, sname='Michel']{Ra\'ul Michel}
\affiliation{Observatorio Astron\'omico Nacional, Instituto de Astronomía, Universidad Nacional Autónoma de México, A.P. 106, 22800 Ensenada, México}
\email{rmm@astro.unam.mx}

\author[orcid=0000-0001-6027-4562,gname=Huifang, sname='Xue']{Huifang Xue}
\affiliation{Department of Physics, Taiyuan Normal University, Jinzhong 030619, People's Republic of China}
\affiliation{Institute of Computational and Applied Physics, Taiyuan Normal University, Jinzhong 030619, People's Republic of China}
\email{hfxue@tynu.edu.cn}

\author[gname=Nian, sname='Liu']{Nian Liu}
\affiliation{School of Physics and Astronomy, China West Normal University, 637009 Nanchong, People's Republic of China}
\email{nian@cwnu.edu.cn}

\author[orcid=0009-0005-2920-7261,gname=Zhongyang, sname='Liu']{Zhongyang Liu}
\affiliation{Institute for Frontiers in Astronomy and Astrophysics, Beijing Normal University, Beijing 102206, People's Republic of China}
\affiliation{School of Physics and Astronomy, Beijing Normal University, Beijing 100875, People's Republic of China}
\email{jerry@mail.bnu.edu.cn}

\author[orcid=0000-0001-7566-9436,gname=Jie, sname='Su']{Jie Su}
\affiliation{Yunnan Observatories, Chinese Academy of Sciences, Kunming 650216, People's Republic of China}
\affiliation{Key Laboratory for the Structure and Evolution of Celestial Objects, Chinese Academy of Sciences, Kunming 650216, People's Republic of China}
\affiliation{International Centre of Supernovae, Yunnan Key Laboratory, Kunming 650216, People's Republic of China}
\email{sujie@ynao.ac.cn}

\author[orcid=0000-0002-0040-8351,gname=Mingfeng, sname='Qin']{Mingfeng Qin}
\affiliation{School of Physics, Nankai University, Tianjin 300071, People's Republic of China}
\email{202331160013@mail.bnu.edu.cn}

\begin{abstract}

A subclass of intermediate mass variables $\delta$ Scuti stars, known as High-amplitude $\delta$ Scuti (HADS) stars, exhibits pronounced radial pulsations. We used ground- and space-based observations to perform an asteroseismic analysis of the HADS star EH~Lib. Following the reduction of light curves, the frequency analysis reveals the dominant frequency as $f_0 =11.3105~\rm{d^{-1}}$ and two newly detected significant frequencies $f_1$ and $f_2$, in addition to the harmonics of $f_0$ and a linear combination. The period change rate is determined as $(1/P_0)(dP_0/dt)=(5.4\pm0.5)\times10^{-9}~\mathrm{yr^{-1}}$ derived from an O-C diagram, which is constructed from 342 times of maximum light spanning over 70 years. Using these observational constraints, along with the metallicities reported in the literature, we construct theoretical models using the stellar evolution code MESA and calculate the theoretical frequencies of the eigenmodes using the oscillation code GYRE. The appropriate models are selected by matching $f_0$, $(1/P_0)(dP_0/dt)$ and luminosity within their respective uncertainties. The results indicate that the observed period change of $\rm{EH~Lib}$ can be attributed to stellar evolutionary effects. The stellar parameters of $\rm{EH~Lib}$ are derived as: the mass of $1.555\pm0.015~M_{\odot}$, the luminosity of $\mathrm{log}\,(L/L_{\odot})=1.26\pm0.02$, and the age of $(1.46\pm0.05)\times10^{9}~\rm{years}$. $\rm{EH~Lib}$ is classified as a single-radial-mode HADS star, located in the subgiant branch, with a helium core and hydrogen-burning shell. This work expands the asteroseismological sample of HADS stars and establishes a foundation for future investigations into their commonalities and specific properties, thereby advancing our understanding of these variables.

\end{abstract}



\section{Introduction}\label{section:1}

Asteroseismology, which emerged nearly three decades ago, has grown to a mature field of the modern astrophysics. It offers a unique opportunity to probe the interiors of pulsating stars, bringing new insights in stellar structure and evolution theories \citep{2021RvMP...93a5001A,2022ARA&A..60...31K}. As a classical type of variables, $\delta$ Scuti stars are intrinsically pulsating variables of spectral types A to F, with masses roughly ranging from 1.5 to 2.5 M$_\odot$. These stars are located at the intersection of the main sequence (MS), pre-MS, or post-MS and the classical instability strip in the Hertzsprung-Russell (HR) diagram and thus considered to be in a stage of central hydrogen or shell hydrogen burning \citep{1979PASP...91....5B,2005JApA...26..249G,2010aste.book.....A,2011A&A...534A.125U}. Their oscillations are driven by the $\kappa$ mechanism in the second partial ionization zone of helium. $\delta$ Scuti stars typically exhibit radial and/or non-radial p modes, as well as mixed modes. Where low-order p modes have pulsation periods ranging from approximately 18 minutes to 8 hours, and regular spacing sequences of high-order p modes are detected in some $\delta$ Scuti stars \citep{2009AIPC.1170..403H,2014MNRAS.439.2078H,2020Natur.581..147B}.

High-amplitude $\delta$ Scuti (HADS) stars represent a subclass of $\delta$ Scuti stars, defined by brightness variations in the \textit{V}-band exceeding 0.1 mag \citep{2002ApJ...576..963T}, though other sources adopt a threshold of 0.3 mag \citep{2000ASPC..210....3B}. They are rare, as \cite{2008PASJ...60..551L} estimated that less than 1 per cent of $\delta$ Scuti stars are HADS stars. Because the definitions rely solely on light curves phenomenologically, the rarity of HADS stars and the physical distinctions between them and their low‐amplitude counterparts remain poorly understood. Nonetheless, HADS stars have similar characteristics to evolved pulsators such as Cepheid variables in the classical instability strip, including asymmetric light curves, period ratios for multi‐mode variables, and period–luminosity (P-L) relations \citep{1996A&A...312..463P,2000ASPC..210..373M}. Most HADS stars exhibit single or double radial pulsation modes, typically the fundamental and first overtone modes \citep{2009PASP..121..251F,2018ApJ...861...96X,2022MNRAS.512.3551D}. However, some HADS stars display three or more radial pulsation modes or, in some cases, even non-radial modes \citep{2020ApJ...904....5X,2022MNRAS.510.1748N,2024A&A...682L...8N}. HADS stars generally have low rotational velocities \citep{2001A&A...366..178R}, though exceptions exist, such as V2367 Cyg, which has an unusually high rotational velocity of approximately 100 km s$^{-1}$ \citep{2012MNRAS.419.3028B}. The linear period change rates, expressed as $(1/P)(dP/dt)$, and derived from long-term observations, are indicators for determining their evolutionary stages \citep{2008AJ....135.1958F}. For $\delta$ Scuti stars, \cite{1998A&A...332..958B} predicted that increases of $(1/P)(dP/dt)$ are from $10^{-10}~\mathrm{yr^{-1}}$ on the main sequence to $10^{-7}~\mathrm{yr^{-1}}$ for more evolved variables. Recent asteroseismic studies have shown that measured period change rates can serve as an additional and powerful constraint when performing successful asteroseismology on HADS stars \citep{2012AJ....144...92Y,2017MNRAS.467.3122N}.

EH Librae $(\alpha_{J2000}=14^{\text{h}}58^{\text{m}}56^{\text{s}},~\delta_{J2000}=-00^{\circ}56'53'')$ was first identified as a variable star by A. N. Vyssotsky, who detected variations in its density using an objective prism plate, as indicated by \cite{1950PASP...62..166C}. Subsequently, \cite{1950PASP...62..166C} took photoelectric photometry observations, determining a preliminary period of 0.08842 days. \cite{1980CoKon..74....1M} suggested that the star exhibits a relatively stable pulsation period, while \cite{1981AcASn..22..279J} proposed a signal of binary motion and a possible period change. \cite{1986PASP...98..651J} observed the star in $uvby-\beta$ filters, estimating a mean effective temperature of $\langle T_{\rm eff} \rangle=7840$ K, and a mean surface gravity of $\langle$log g$\rangle=4.08~\rm{dex}$. Through a spectroscopic survey, \cite{1997A&AS..122..131S} derived an effective temperature of $T_{\rm eff}=7711~\rm K$ using the calibration method proposed by \cite{1994MNRAS.268..119B}, along with a projected rotational velocity of $\textit{v}\,\textnormal{sin}\,\textit{i}=13.30~\textnormal{km s}^{-1}$. With high-resolution spectroscopy, \cite{2017MNRAS.470.4408K} determined the effective temperature of $T_{\rm eff}=7300\pm100~\rm{K}$, the surface gravity of $\rm{log~g}=3.9\pm0.1~\rm{dex}$, the projected rotational
velocity of $\textit{v}\,\textnormal{sin}\,\textit{i}=15\pm1$ km s$^{-1}$, and the iron abundance of $\textnormal{log}\,\epsilon\,\textnormal{(Fe)}=7.11\pm0.12~\rm{dex}$. Furthermore, \cite{2017IBVS.6231....1P} monitored the star between 2013 and 2016. By combining these observations with previously compiled times of maximum light, they confirmed the existence of a measurable period change rate.

Alias frequencies arising from almost regular gaps in ground-based observations hinder accurate frequency analysis. However, the continuous, high-precision, and long-duration photometric data from the Transiting Exoplanet Survey Satellite (TESS) mission \citep{2014SPIE.9143E..20R,2015JATIS...1a4003R} , which performs a near all-sky survey, allows significant advancement in determining the pulsation of $\rm{EH~Lib}$. When combined with a newly determined linear period change rate from recent observations, these data provide a valuable opportunity to place tighter constraints on its evolutionary state. Accordingly, our current study aims to determine the stellar structure and evolutionary stage of $\rm{EH~Lib}$, and construct its asteroseismological models for the first time.

The structure of the paper is as follows. Section~\ref{section:2} presents new photometric observations and associated data reduction procedures. Section~\ref{section:3} focuses on the frequency analysis, mode identification and determination of the period change rate of $\rm{EH~Lib}$. Section~\ref{section:4} describes the construction of stellar evolutionary tracks and the identification of the optimal models. Section~\ref{section:5} and \ref{section:6} provide a discussion and conclusions of the study, respectively.

\section{Observations and data reduction}\label{section:2}

In this study, ground-based photometric data collected by our team from 2009 to 2018, along with high-precision space-based observations collected during TESS Sector 51, are utilized.

\subsection{Ground-based Photometry}
\begin{figure}
    \centering
    \includegraphics[width=\columnwidth]{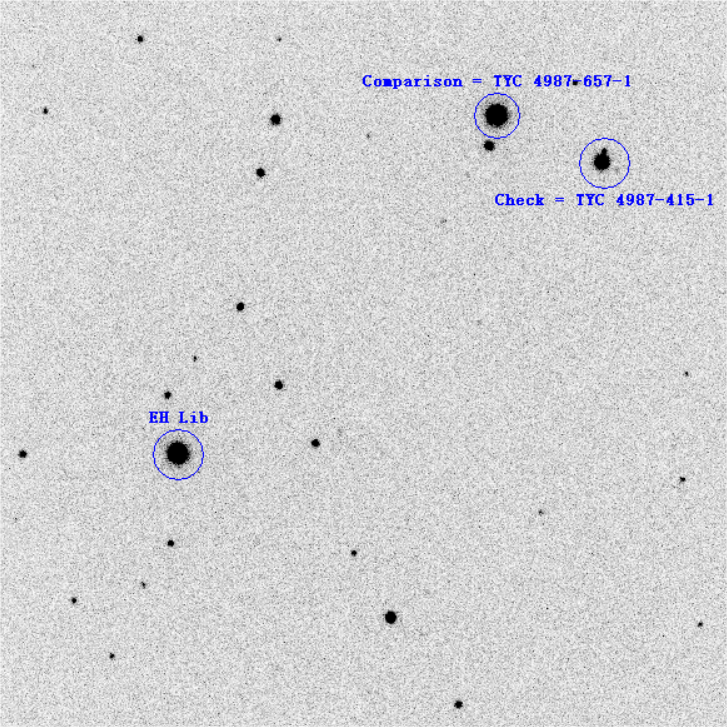}
    \caption{A CCD image of $\rm{EH~Lib}$ taken with the Xinglong 2.16-m Telescope. The field of view is $9^{\prime}$.36 $\times$ $9^{\prime}$.36. $\rm{EH~Lib}$, the comparison star (TYC 4987-657-1) and the check star (TYC 4987-415-1) are marked. South is up and East is to the left.}
    \label{fig:Figure1}
\end{figure}

Time-series photometric observations for $\rm{EH~Lib}$ were taken at the Yunnan Astronomical Observatory of China, Xinglong station of National Astronomical Observatories of China, and National Astronomical Observatory San Pedro Mártir of Mexico between 2009 and 2018. Figure~\ref{fig:Figure1} presents an image of $\rm{EH~Lib}$ taken with the Xinglong 2.16-m Telescope, with $\rm{EH~Lib}$, the comparison star, and the check star marked. The details of these three stars from SIMBAD\footnote{\url{https://simbad.u-strasbg.fr/simbad/}} \citep{2000A&AS..143....9W} are listed in Table~\ref{tab:Tabels2}. A total of 11694 frames were obtained across 55 nights of observations. Table~\ref{tab:Tabel1} lists the journal of observations.

\begin{table*}
    \centering
    \renewcommand{\arraystretch}{1.5}
    \caption{The comparison star and the check star used in the photometry of $\rm{EH~Lib}$.}
    \label{tab:Tabels2}
    \begin{tabular}{p{120pt}p{60pt}p{60pt}ccc}
        \hline
        Star name & $\alpha~[J2000]$ & $\delta~[J2000]$ & \textit{V} & \textit{B} & \textit{B - V} \\
        \hline
        Object = EH Librae & $14^\text{h}58^\text{m}55^\text{s}.92$ & $-00^\circ56'53''.01$ & 9.83 & 10.09 & 0.26 \\
        Comparison = TYC 4987-657-1 & $14^\text{h}58^\text{m}39^\text{s}.80$ & $-01^\circ01'17''.63$ & 9.80 & 10.85 & 1.05 \\
        Check = TYC 4987-415-1 & $14^\text{h}58^\text{m}34^\text{s}.25$ & $-01^\circ00'42''.55$ & 11.28 & 12.52 & 1.24 \\
        \hline
    \end{tabular}
\end{table*}

\begin{table*}
    \centering
    \renewcommand{\arraystretch}{1.5}
    \caption{Journal of the photometric observations for $\rm{EH~Lib}$. YAO=Yunnan Astronomical Observatory of China, XL=Xinglong station of National Astronomical Observatories of China, SPM=National Astronomical Observatory San Pedro Mártir of Mexico. Telescope=Telescope aperture in cm. Filter symbols: \textit{V}, \textit{R}=Johnson \textit{V}, \textit{R}.}
    \label{tab:Tabel1}
    \begin{tabular}{r c c c c c c c}
        \hline
        \multicolumn{1}{c}{Date} & Observatory & Telescope & CCD & Pixel & Filter & Night & Frames \\
        \hline
        2009, Feb. & YAO & 101.6 & Andor DW436 & 1024$\times$1024 & \textit{V} & 10 & 1254 \\
        2014, Feb. & YAO & 101.6 & Andor DW436 & 2048$\times$2048 & \textit{R} & 6 & 253 \\
        & & & & & \textit{V} & 6 & 254 \\
        Mar. & SPM & 84 & E2V-4240 & 1076$\times$1024 & \textit{R} & 6 & 805 \\
        & & & & & \textit{V} & 6 & 799 \\
        & XL & 85 & PI BFT512 & 1024$\times$1024 & \textit{R} & 6 & 1415 \\
        & & & & & \textit{V} & 6 & 1425 \\
        Apr. & XL & 85 & PI BFT512 & 1024$\times$1024 & \textit{R} & 1 & 250 \\
        & & & & & \textit{V} & 1 & 250 \\
        2015, Jan. & YAO & 101.6 & Andor DW436 & 1024$\times$1024 & \textit{R} & 6 & 365 \\
        & & & & & \textit{V} & 9 & 608 \\
        & SPM & 84 & E2V-4240 & 1076$\times$1024 & \textit{V} & 1 & 222 \\
        Mar. & XL & 85 & Andor DZ936 & 2048$\times$2048 & \textit{R} & 1 & 317 \\
        & & & & & \textit{V} & 1 & 317 \\
        Apr. & SPM & 84 & E2V-4240 & 1076$\times$1024 & \textit{R} & 4 & 219 \\
        & & & & & \textit{V} & 4 & 215 \\
        2016, Feb. & YAO & 101.6 & Andor DW436 & 1024$\times$1024 & \textit{R} & 4 & 355 \\
        & & & & & \textit{V} & 4 & 358 \\
        2017, Jan. & SPM & 84 & E2V-4240 & 1036$\times$1024 & \textit{R} & 4 & 288 \\
        & & & & & \textit{V} & 4 & 282 \\
        2018, Mar. & XL & 216 & E2V CCD42-40 & 2048$\times$2048 & \textit{R} & 3 & 1443 \\
        \hline
    \end{tabular}
\end{table*}

The observed frames are first corrected by subtracting the master bias and dark frames, and then divided by the master flat frame following the standard CCD data reduction procedure. Aperture and differential photometry are subsequently performed on $\rm{EH~Lib}$, the comparison star, and the check star. This procedure is executed using the package IRAF \citep{1986SPIE..627..733T,1993ASPC...52..173T} and the software AstroImageJ \citep{2017AJ....153...77C}. The magnitude differences between $\rm{EH~Lib}$ and the comparison star are calculated to determine the relative magnitudes of $\rm{EH~Lib}$. Those between the check star and the comparison star are used to estimate the photometric precisions from the standard deviations of the differential magnitudes.

\begin{figure}
    \centering
    \includegraphics[width=\columnwidth]{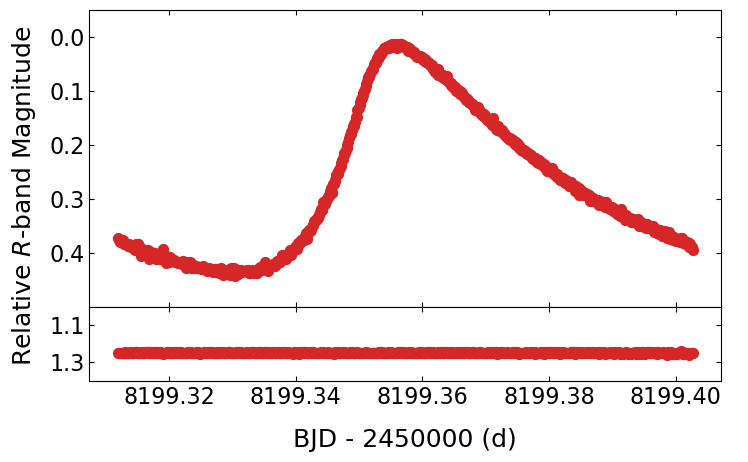}
    \caption{An observed light curve of $\rm{EH~Lib}$ in \textit{R}-band, taken with the Xinglong 2.16-m Telescope on 21 March 2018. The top panel shows the magnitude differences between $\rm{EH~Lib}$ and the comparison star, and the bottom panel shows the magnitude differences between the check star and the comparison star.}
    \label{fig:Figure2}
\end{figure}

After data reduction, the whole light curves in \textit{V} and \textit{R} bands are presented in Figures~\ref{fig:FigureA1} and \ref{fig:FigureA2}, respectively. Figure~\ref{fig:Figure2} shows an example of these observed light curves.

\subsection{Space-based Photometry}
\begin{figure*}
    \centering
    \includegraphics[width=\textwidth]{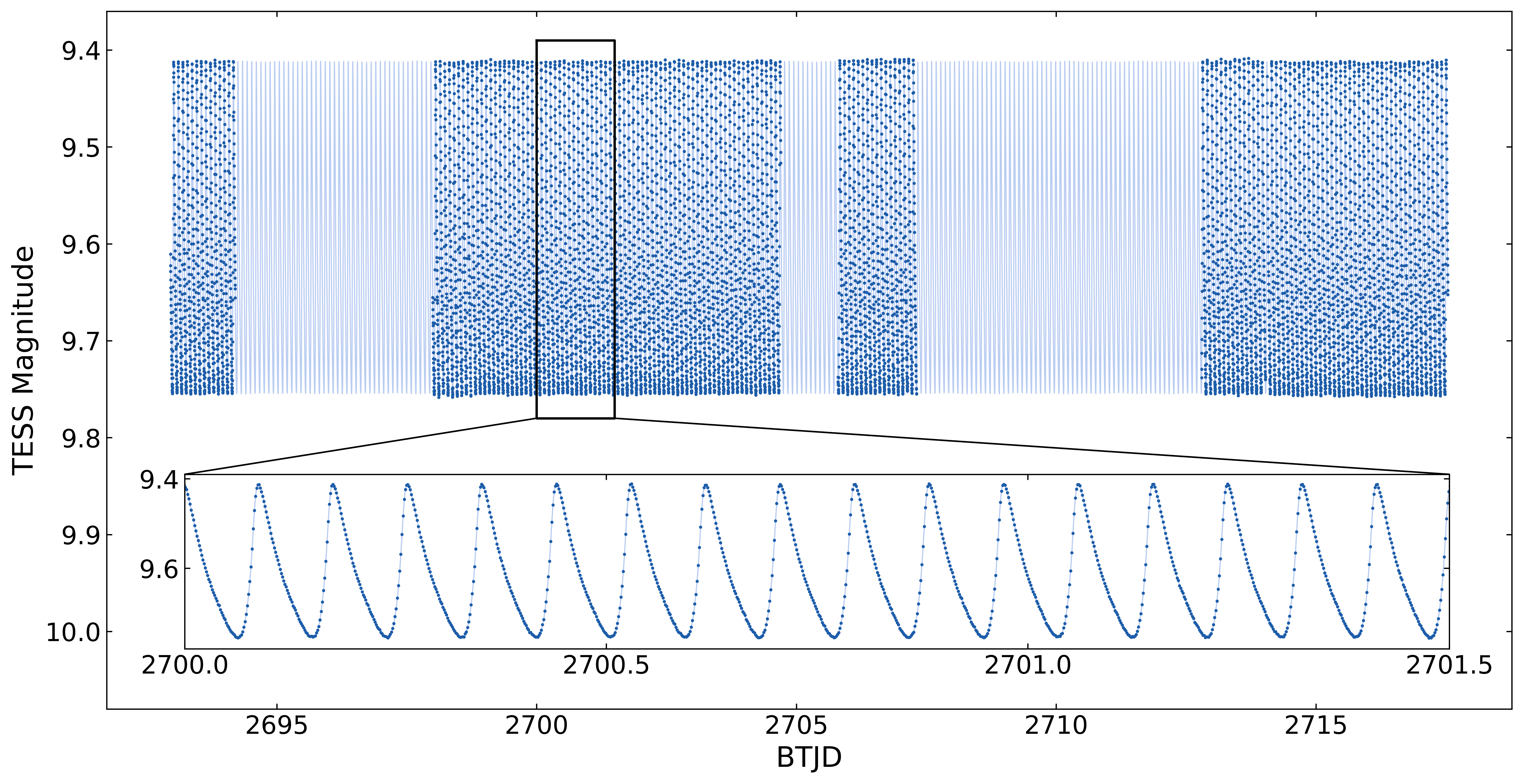}
    \caption{Blue points present all the reduced data points observed by TESS, while the light blue line presents the fitted curve obtained from the extracted frequency solution listed in Table~\ref{tab:Tabel2}. The small panel shows the zoomed-in view of a part of the light curves.}
    \label{fig:Figure3}
\end{figure*}

TESS observed $\rm{EH~Lib}$ (TIC 157861023) with a 2-min cadence during Sector 51. All target pixel files (TPFs) are downloaded from the Mikulski Archive for Space Telescopes (MAST)\footnote{\url{https://archive.stsci.edu}} using the \textmd {Lightkurve} package \citep{2018ascl.soft12013L,2021zndo...1181928B}. Aperture sizes are carefully selected on the TPFs to produce optimal light curves. After removing outliers and converting the normalized flux to TESS magnitude, a quartic polynomial function is applied to fit and then detrend the light curves, resulting in a total of 10,795 reduced data points. The final TESS light curves are presented in Figure~\ref{fig:Figure3}.

\section{Pulsation Analysis}\label{section:3}
\subsection{Frequency Analysis}\label{section:3.1}
Given that the frames observed by TESS provide the most continuous, high-precision, and long-duration data, frequency analysis is performed exclusively on the TESS light curves. The software Period04 \citep{2005CoAst.146...53L} computes discrete Fourier transformations of the light curves to search for significant peaks in the amplitude spectra below Nyquist frequency. In this process, a Hamming window is applied to reduce sidelobes, particularly the first one, thereby enhancing the clarity of the main peaks, and to suppress spectral leakage caused by data gaps and irregular sampling in TESS observations \citep{1978IEEEP..66...51H}. The light curves are subsequently fitted based on the extracted frequencies using a least-squares approach with the following formula,

\begin{equation}
    m = m_0 + \sum_i A_i \sin \big( 2\pi \big( f_i t + \phi_i \big) \big)
    \label{eq:equation1}
\end{equation}

\begin{figure}
    \centering
    \includegraphics[width=\columnwidth]{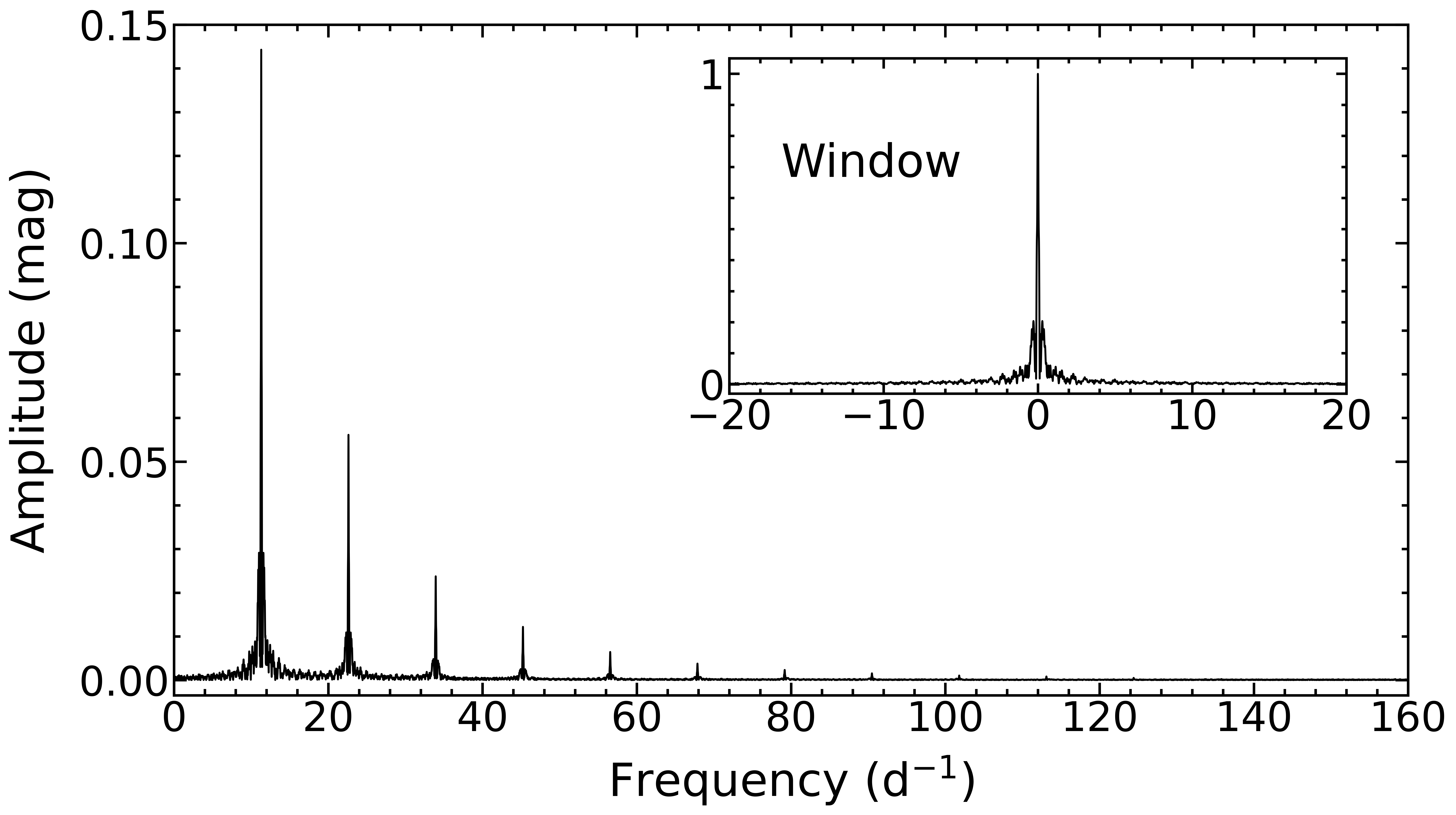}
    \caption{The Fourier amplitude spectrum of the TESS light curves. The corresponding spectral window is shown in the inset.}
    \label{fig:Figure4}
\end{figure}

\begin{figure*}
    \centering
    \includegraphics[width=\textwidth]{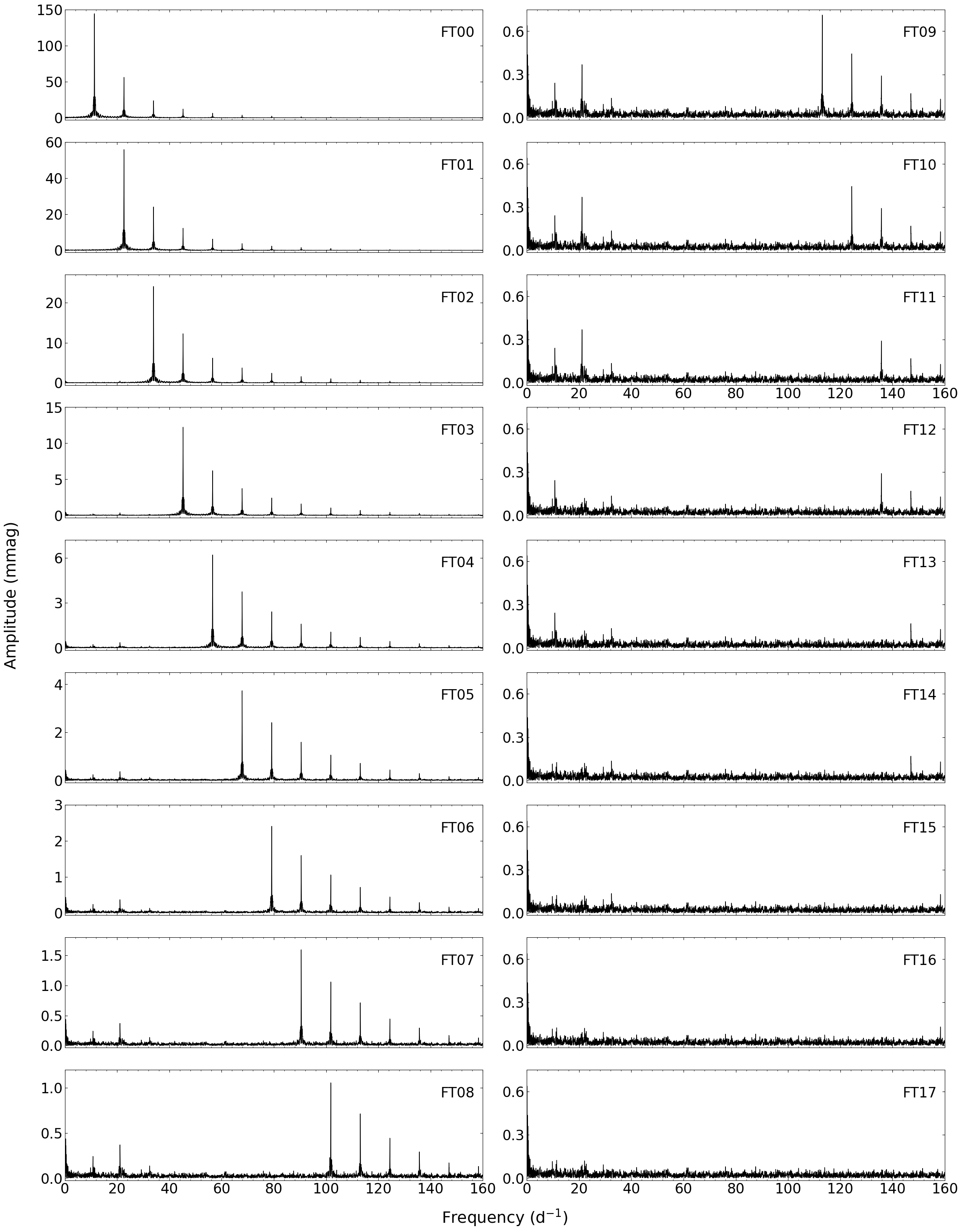}
    \caption{The Fourier amplitude spectra of the TESS light curves with the frequency pre-whitening process shown.}
    \label{fig:Figure5}
\end{figure*}

Table~\ref{tab:Tabel2} lists the frequency solution, including the dominant frequency $f_0 =11.3105~\rm{d^{-1}}$ and other frequencies, each with a signal-to-noise (S/N) ratio exceeding 5.3 according to the criterion of \cite{2021AcA....71..113B}. It should be noted that the frequencies $f_1$ and $f_2$ are found for the first time. The Rayleigh resolution is 0.061 $\rm{d^{-1}}$. Figure~\ref{fig:Figure4} displays the amplitude spectrum with its corresponding spectral window of the TESS light curves, and Figure~\ref{fig:Figure3} presents the corresponding fitting curves shown in light blue lines based on the solution. Figure~\ref{fig:Figure5} shows the amplitude spectra of the frequency pre-whitening process. Since the low-frequency domain (0-2 $\rm{d^{-1}}$) of the amplitude spectra is affected by the systemic noise, such as the spacecraft motions, the instrument sensitivity instability and scattered light contamination, the peaks in this domain are excluded from the analysis.

\begin{table}
    \centering
    \renewcommand{\arraystretch}{1.2}
    \caption{The frequency solution of the light curves observed by TESS for $\rm{EH~Lib}$.}
    \label{tab:Tabel2}
    \begin{tabular}{l r c c r}
        \hline
        \multicolumn{1}{c}{No.} & \multicolumn{1}{c}{ID} & Frequency ($\rm{d^{-1}}$) & \multicolumn{1}{c}{Amplitude (mmag)} & \multicolumn{1}{c}{S/N} \\
        \hline
        F1 & $f_0$ & 11.310514(2) & 144.01 & 4351.5 \\
        F2 & $2f_0$ & 22.62100(1) & 55.96 & 1937.6 \\
        F3 & $3f_0$ & 33.93152(8) & 24.01 & 948.8 \\
        F4 & $4f_0$ & 45.24213(0) & 12.23 & 497.2 \\
        F5 & $5f_0$ & 56.5525(9) & 6.19 & 294.1 \\
        F6 & $6f_0$ & 67.8630(1) & 3.73 & 175.4 \\
        F7 & $7f_0$ & 79.1736(0) & 2.40 & 101.8 \\
        F8 & $8f_0$ & 90.4842(3) & 1.59 & 70.1 \\
        F9 & $9f_0$ & 101.7949(0) & 1.06 & 46.3 \\
        F10 & $10f_0$ & 113.104(5) & 0.71 & 32.2 \\
        F11 & $11f_0$ & 124.415(2) & 0.44 & 21.6 \\
        F12 & $f_1$ & 21.107(0) & 0.37 & 13.6 \\
        F13 & $12f_0$ & 135.726(7) & 0.29 & 12.6 \\
        F14 & $f_2$ & 10.750(0) & 0.24 & 8.0 \\
        F15 & $13f_0$ & 147.041(6) & 0.17 & 7.8 \\
        F16 & $f_0+f_1$ & 32.419(7) & 0.14 & 6.3\\
        F17 & $14f_0$ & 158.348(8) & 0.13 & 6.0 \\
        \hline
    \end{tabular}
\end{table}

Candidate linear combinations of the extracted frequencies, expressed as $f_a \pm \sigma_a=f_b \pm \sigma_b \pm f_c \pm \sigma_c$, are evaluated using the criterion of \cite{2013MNRAS.429.1585F},  
namely $\sigma_a \leq 3 \times (\sigma_b+\sigma_c)$. When this criterion is satisfied, the component with the lowest amplitude is identified as a combination, while the higher ones as independent frequencies which are $f_0$ and $f_1$ for $\rm{EH~Lib}$. There are no detectable combinations of $f_2$. Candidate harmonics of $f_0$ are identified following the similar criterion.

\subsection{Mode Identification}\label{section:3.2}
Among the three frequencies $f_0$, $f_1$ and $f_2$ of $\rm{EH~Lib}$ listed in Table~\ref{tab:Tabel2}, the amplitude of $f_0$ is significantly larger than those of $f_1$ and $f_2$, the corresponding period $P_0=0.0884133\pm0.0000001~\mathrm{days}$ of $f_0$ is the pulsation period $P$ of $\rm{EH~Lib}$. Based on the typical pulsational characteristics of HADS stars, $f_0$ is assumed to be a radial mode, while $f_1$ and $f_2$ are not. Consequently, once its mode is identified, only $f_0 = 11.310514 \pm 0.000003~\rm{d^{-1}}$ is used to constrain the stellar models.

Traditionally, pulsation modes can be identified by calculating pulsation constants \citep{1975ApJ...200..343B}. The definition is,

\begin{align}
    Q = P \sqrt{ \frac{ \bar{\rho} }{ \bar{\rho}_{\odot} } }\label{eq:eq1_1}
\end{align}

Where $P$ is in days, $\bar{\rho}$ the mean stellar density, and $\bar{\rho}_{\odot}$ the mean solar density. Following \cite{1990DSSN....2...13B}, Equation~\eqref{eq:eq1_1} can be expressed as,

\begin{align}
    \log Q = \log P+ 0.5 \log g + 0.1 M_{\text{bol}} + \log T_{\text{eff}} - 6.456\label{eq:eq1_2}
\end{align}

Here, $M_{\mathrm{bol}}$ is the absolute bolometric magnitude. Together with $M_{\mathrm{bol}}$, the bolometric luminosity $\mathrm{log}\,(L/L_{\odot})$ of $\rm{EH~Lib}$ is derived from its parallax, $\varpi=2.7339\pm0.0192~\rm{mas}$, provided by Gaia DR3 \citep{2023A&A...674A...1G}, using the following equations,

\begin{align}
    &M_{\mathrm{bol}}=m_G+BC_G-A_G+5\rm{log}\varpi+5\label{eq:equation13}\\
    &-2.5\mathrm{log}\,(L/L_{\odot})=M_{\mathrm{bol}}-M_{\mathrm{bol},\odot}\label{eq:equation14}
\end{align}

The calculations incorporate the mean G-band magnitude $m_G=9.8853\pm0.0058$, the interstellar extinction $A_G=0.1299$, and the bolometric correction $BC_G=0.019$, estimated following the methods proposed by \cite{2010A&A...523A..48J} and \cite{2018A&A...616A...8A}. Adopting the absolute bolometric magnitude of the Sun $M_{\mathrm{bol},\odot}=4.74$ as defined by IAU \citep{2015arXiv151006262M}, the absolute bolometric magnitude of $\rm{EH~Lib}$ is determined as $M_{\mathrm{bol}}=1.959\pm0.064$, corresponding to a bolometric luminosity of $\mathrm{log}\,(L/L_{\odot})=1.11\pm0.03$.

Substituting $P=0.0884133\pm0.0000001~\mathrm{days}$, $M_{\mathrm{bol}}=1.959\pm0.064$, the effective temperature $T_{\rm eff}=7300\pm100~\rm{K}$ and the surface gravity $\rm{log~g}=3.9\pm0.1~\rm{dex}$ from \cite{2017MNRAS.470.4408K} into Equation~\eqref{eq:eq1_2} yields a pulsation constant for $\rm{EH~Lib}$ of $Q=0.032\pm0.001~\mathrm{days}$. According to \cite{1975ApJ...200..343B}, fundamental radial modes of $\delta$ Scuti stars typically have $Q>0.029~\mathrm{days}$, thus $f_0$ satisfies this criterion.

Besides, the pulsation mode can be identified by comparing the \textit{V}-band absolute magnitude $M_v$ and period $P_0$ with the newly calibrated P-L relations for $\delta$ Scuti stars \citep{2019MNRAS.486.4348Z,2020MNRAS.493.4186J,2021PASP..133h4201P,2022MNRAS.516.2080B}. For instance, \cite{2022MNRAS.516.2080B} derived the following P–L relation for $\delta$ Scuti stars pulsating in the fundamental radial mode,

\begin{equation}
    Mv=(-3.01\pm0.07)~\mathrm{log}(P_0/d)-(1.40\pm0.07)
    \label{eq:eq1_3}
\end{equation}

Analogously following the methods of \cite{2022MNRAS.516.2080B}, the \textit{V}-band apparent magnitude $m_v=9.88$ of $\rm{EH~Lib}$ is obtained from ASAS-SN\footnote{\url{https://asas-sn.osu.edu/variables}}, and the corresponding extinction coefficient $A_v=0.176$ is calculated utilizing the dust-maps Python package of \cite{2019ApJ...887...93G}. The \textit{V}-band absolute magnitude $M_v$ is then calculated by the below equation,

\begin{equation}
    M_v=m_v+5\rm{log}\varpi+5-A_v
    \label{eq:equation6}
\end{equation}

Thus, the \textit{V}-band absolute magnitude of $\rm{EH~Lib}$ is determined as $M_v=1.888$. According to Equation~\eqref{eq:eq1_3}, a HADS star pulsating in the fundamental radial mode with period $P_0$ should have $1.627 \leq M_v \leq 1.915$. Moreover, $M_v=1.888$ is consistent with the corresponding P–L relations of \cite{2019MNRAS.486.4348Z,2020MNRAS.493.4186J,2021PASP..133h4201P}.

In one word, $f_0$ is identified as the fundamental mode of the radial pulsations of $\rm{EH~Lib}$.

\subsection{Period Change Rate}
Constraining theoretical models for HADS stars is challenging with only one frequency. Hence, an additional pulsation parameter, period change rate, can be helpful. Typically, the classical O−C method \citep{2005ASPC..335....3S} can be used to derive the linear period change rate of single-radial-mode HADS stars. Based on the new observations, light curves around the light maxima of EH Lib are fitted using quartic polynomials to determine the times of maximum light. The uncertainties of these times of maximum light are estimated using the Markov Chain Monte Carlo (MCMC) method with 20,000 iterations performed for each maximum. Typical uncertainties are found to be less than 0.0003 days. Table~\ref{tab:Tabel4} lists 53 newly determined times of maximum light derived from the ground-based \textit{V}-band observations. The same analysis is applied to the ground-based \textit{R}-band observations and the TESS data, yielding 38 and 158 newly determined times of maximum light, respectively, as listed in Tables~\ref{tab:TabelA1} and~\ref{tab:TabelA2}.

\begin{figure*}
    \centering
    \includegraphics[width=\textwidth]{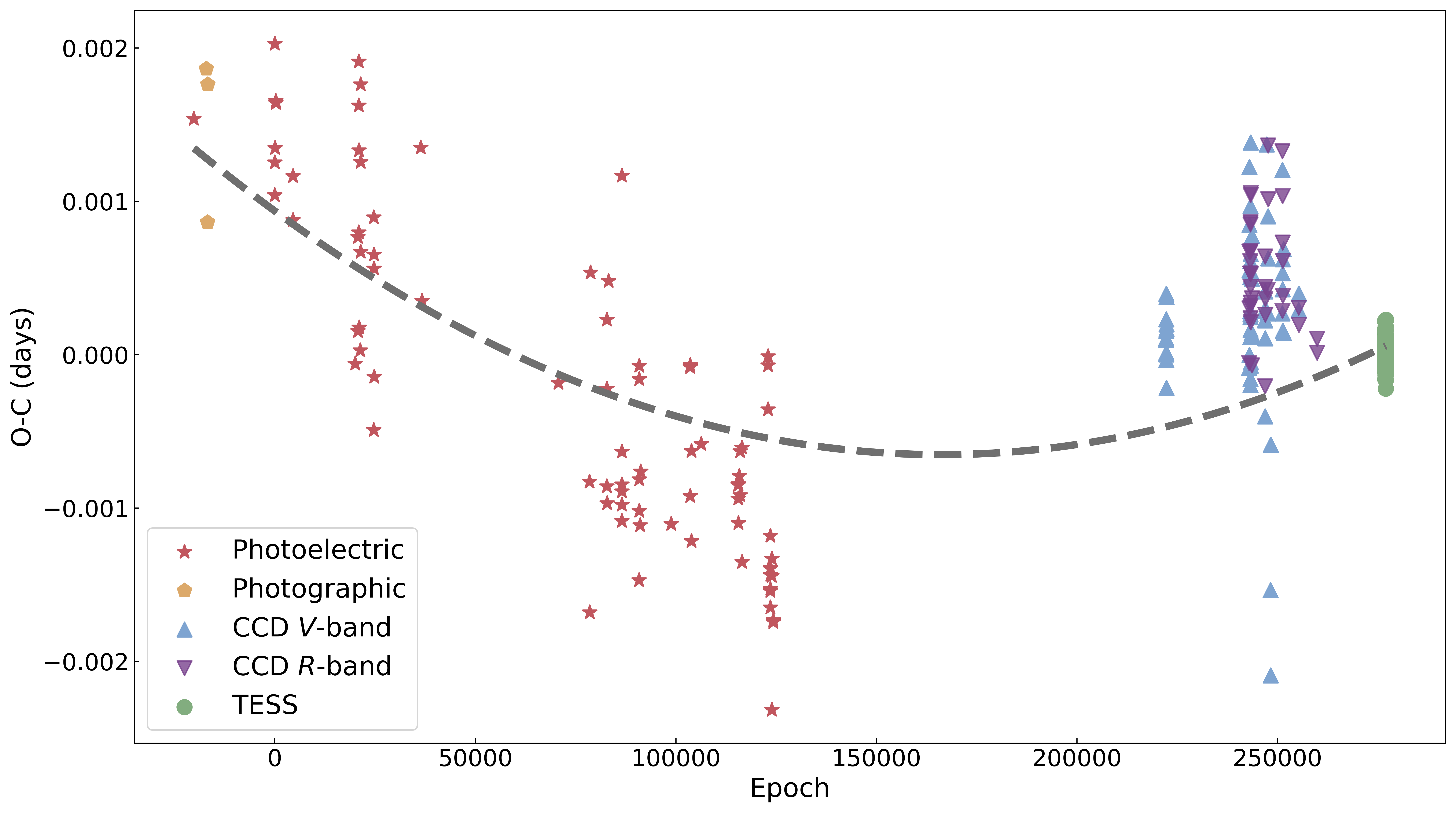}
    \caption{O−C diagram for $\rm{EH~Lib}$ calculated using the ephemeris from Equation~\eqref{eq:equation7}, with the parabolic polynomial fit represented by a dashed curve. Different markers indicate data obtained from various detector types: red stars for photoelectric, orange pentagons for photographic, blue up-pointing triangles for CCD \textit{V}-band, purple down-pointing triangles for CCD \textit{R}-band and green circles for TESS data, as described in the labels.}
    \label{fig:Figure6}
\end{figure*}

\begin{table}
\centering
\renewcommand{\arraystretch}{1.2}
\setlength{\tabcolsep}{4pt}
\caption{Times of maximum light of $\rm{EH~Lib}$ from new ground-based \textit{V}-band observations. $T_{\rm{max}}$ is in BJD-2450000, O-C is in days. Epoch and O−C are derived from the ephemeris formula in Equation~\eqref{eq:equation7}.}
\label{tab:Tabel4}
\begin{tabular}{ccrccr}
\hline
\textbf{$T_{\rm{max}}$} & Epoch & \multicolumn{1}{c}{O-C} & \textbf{$T_{\rm{max}}$} & Epoch & \multicolumn{1}{c}{O-C} \\
\hline
4867.41847 & 222180 & 0.00011 & 6732.31957 & 243273 & 0.00016  \\
4868.39101 & 222191 & 0.00010 & 6733.02737 & 243281 & 0.00066  \\
4869.36346 & 222202 & 0.00000 & 6734.97318 & 243303 & 0.00138  \\
4872.45792 & 222237 & 0.00000 & 6736.29793 & 243318 & -0.00007  \\
4873.34221 & 222247 & 0.00016 & 6736.38696 & 243319 & 0.00055  \\
4873.43086 & 222248 & 0.00040 & 6737.27050 & 243329 & -0.00005  \\
4874.31456 & 222258 & -0.00003 & 6761.23103 & 243600 & 0.00049  \\
4874.40302 & 222259 & 0.00001 & 6761.31973 & 243601 & 0.00077  \\
4875.28737 & 222269 & 0.00023 & 7049.45802 & 246860 & 0.00022  \\
4877.32085 & 222292 & 0.00020 & 7050.42994 & 246871 & -0.00040  \\
4878.29298 & 222303 & -0.00022 & 7052.46407 & 246894 & 0.00023  \\
4878.38198 & 222304 & 0.00038 & 7053.43668 & 246905 & 0.00029  \\
4879.35432 & 222315 & 0.00017 & 7054.40935 & 246916 & 0.00041  \\
6703.40818 & 242946 & -0.00008 & 7056.44274 & 246939 & 0.00030  \\
6704.38081 & 242957 & 0.00000 & 7057.41509 & 246950 & 0.00011  \\
6705.35458 & 242968 & 0.00122 & 7106.30779 & 247503 & 0.00027  \\
6705.44232 & 242969 & 0.00055 & 7117.80215 & 247633 & 0.00090  \\
6708.36025 & 243002 & 0.00085 & 7119.83538 & 247656 & 0.00063  \\
6723.92026 & 243178 & 0.00012 & 7435.29449 & 251224 & 0.00120  \\
6724.00917 & 243179 & 0.00061 & 7435.38197 & 251225 & 0.00027  \\
6725.95393 & 243201 & 0.00028 & 7437.32722 & 251247 & 0.00043  \\
6726.04257 & 243202 & 0.00051 & 7440.33338 & 251281 & 0.00053  \\
6729.31362 & 243239 & 0.00026 & 7441.30601 & 251292 & 0.00062  \\
6729.40201 & 243240 & 0.00024 & 7441.39396 & 251293 & 0.00016  \\
6730.37411 & 243251 & -0.00020 & 7795.04716 & 255293 & 0.00029  \\
6730.99417 & 243258 & 0.00097 & 7798.93745 & 255337 & 0.00040  \\
6731.34670 & 243262 & -0.00016 &  &  &   \\
\hline
\end{tabular}
\end{table}

For the O−C analysis, 93 times of maximum light reported in previous literature \citep{1950PASP...62..166C,1952AJ.....57Q..64A,1957AJ.....62..108F,1961Obs....81..199S,1966CoLPL...5....3F,1966BAN....18..387O,1977IBVS.1310....1K,1979AAS...36...51G,1980CoKon..74....1M,1981AcASn..22..279J,1985IBVS.2810....1H,1986PASP...98..651J,1992IBVS.3769....1Y,2017IBVS.6231....1P}, as listed in Table~\ref{tab:TabelA3}, are combined with the newly determined values. Since the uncertainties for the times of maximum light reported in the literature were not provided, weights are assigned according to the detector type: 0.5 for photographic, 0.9 for photoelectric, 1.0 for CCD \textit{V}-band, and 2.0 for TESS data. Additionally, for a given pulsation period, there is a systematic offset between the O−C values derived from \textit{V}-band and \textit{R}-band observations, with a mean difference of 0.00011 days and a standard deviation of 0.00030 days. To account for this discrepancy, a weight of 0.1 is assigned to CCD \textit{R}-band data. Furthermore, to correct for the effects of leap seconds, motion, and related factors, all time stamps used in this study are converted to Barycentric Julian Date (BJD) in Barycentric Dynamical Time (TDB). Incorporating these weights, along with the pulsation period determined from TESS observations, the data are used to derive the linear ephemeris formula,

\begin{equation}
    C = \rm{BJD}~2435223.7587 + 0.088413267 \times E
    \label{eq:equation7}
\end{equation}

Figure~\ref{fig:Figure6} presents the O−C diagram, which exhibits long-term variability. As a result, a parabolic polynomial is used to fit the OC values, as indicated by the dashed curve. Incorporating the assigned weights, the parabolic term is determined, yielding the following parabolic ephemeris,

\begin{equation}
    C = \rm{BJD}~2435223.7596 + 0.088413248 \times E + 5.76 \times 10^{-14}~E^2
    \label{eq:equation8}
\end{equation}

The residuals of the fit yield a root-mean-square (RMS) value of 0.00057 days. The parabolic coefficient deviates from zero by 11.34~$\sigma$, demonstrating the necessity of the parabolic fit.

Considering that $\rm{EH~Lib}$ is a single-radial-mode HADS star, whose pulsation is dominated by $f_0$, the period change rate of $f_0$ is consistent with that of the star. Based on the parabolic term in Equation~\eqref{eq:equation8}, the period change rate of $f_0$ can be derived as follows,

\begin{equation}
    \frac{1}{P_0}\frac{dP_0}{dt}=-\frac{1}{f_0}\frac{df_0}{dt}=(5.4\pm0.5)\times10^{-9}~\mathrm{yr^{-1}}
    \label{eq:equation9}
\end{equation}

\section{Theoretical Modeling}\label{section:4}
In this section, the theoretical models of $\rm{EH~Lib}$ are constructed to constrain the stellar parameters of the star, using the open-knowledge software instrument Modules for Experiments in Stellar Astrophysics (MESA, version r22.11.1; \citealt{2011ApJS..192....3P,2013ApJS..208....4P,2015ApJS..220...15P,2018ApJS..234...34P,2019ApJS..243...10P,2023ApJS..265...15J}). The one-dimensional stellar evolution module, MESA star, is used to compute stellar evolutionary tracks for different initial input parameters. Furthermore, the stellar oscillation code, GYRE (version v6.0; \citealt{2013MNRAS.435.3406T,2018MNRAS.475..879T,2020ApJ...899..116G}) coupled with MESA, is used to calculate the frequencies of the eigenmodes at each step along the evolutionary tracks.

\subsection{Physical Parameters}
The high-resolution spectra of $\rm{EH~Lib}$, obtained by \cite{2017MNRAS.470.4408K} using the ARC Echelle Spectrograph (ARCES) mounted at the 3.5-m telescope of the Apache Point Observatory (USA), provides the basis for determination of stellar parameters of this star. The spectra cover a wavelength range of 3200-10000 Å with a resolving power of $R\sim31500$. Based on the reported iron abundance from \cite{2017MNRAS.470.4408K}, the metallicity of $\rm{EH~Lib}$ is estimated as $\mathrm{[Fe/H]}=-0.39\pm0.12$, which is adopted for subsequent calculations.

The initial metal abundance $Z$ is calculated using the approximation proposed by \cite{2012MNRAS.427..127B}, as described in \cite{2021MNRAS.506.6117W},

\begin{align}
    &\mathrm{[M/H]}=\,\mathrm{log}\,(Z/X)-\mathrm{log}\,(Z/X)_{\odot}\label{eq:equation10}\\
    &Y=0.2485+1.78Z\label{eq:equation11}\\
    &X+Y+Z=1\label{eq:equation12}
\end{align}

The spectroscopically determined $\mathrm{[Fe/H]}$ is utilized as a proxy for the total metallicity $\mathrm{[M/H]}$. This treatment assumes that the alpha-element to iron ratio, $\mathrm{[\alpha/Fe]}$, is approximately zero. This assumption is justified for $\rm{EH~Lib}$, a young A7-type star located in the Galactic disk. Large spectroscopic surveys of A, F and G stars in the solar neighbourhood have shown that disk stars with near-solar metallicities typically exhibit $\mathrm{[\alpha/Fe]}$ values close to zero \citep{2011A&A...530A.138C,2014A&A...562A..71B}. Therefore, in the absence of detailed individual abundances for EH Lib, adopting $\mathrm{[M/H]} \approx \mathrm{[Fe/H]}$ is a reasonable approximation for the stellar evolution calculations.

Here, $(Z/X)_{\odot}=0.0207$ \citep{2012MNRAS.427..127B} and $Y$ represents the helium abundance. The metal abundance is estimated as $Z \in [0.0047,0.0081]$.

\subsection{Setup of the Model Calculation}
Since $\rm{EH~Lib}$ is classified as a HADS star, stellar evolutionary tracks are constructed for masses ranging from 1.50 $M_{\odot}$ to 2.50 $M_{\odot}$ with the step size of 0.01 $M_{\odot}$. Three chemical compositions are adopted as $(X,Y,Z)=($$0.7384,0.2569,0.0047),(0.7343,0.2595,0.0062)$ and $(0.7290,0.2629,0.0081)$. The inlist files are available\footnote{\url{https://github.com/Xiran-Xie/EHLib-inlist-files}}.

The default opacity tables provided with MESA, which adopt the solar chemical abundance mixture of \cite{1998SSRv...85..161G}, are used \citep{2011ApJS..192....3P}. They are constructed by combining OPAL Type 1 and 2 tables \citep{1993ApJ...412..752I,1996ApJ...464..943I} for the high-temperature regime, tables from \cite{2005ApJ...623..585F} for the low temperature regime, and tables from OP \citep{2005MNRAS.362L...1S} can be used in place of OPAL as the table format is identical. In regions of overlapping coverage a smooth blend is applied. OPAL-based tables are adopted here because \citet{2023ApJ...942L..38D} reported that they give better agreement with the observed effective temperatures and luminosities of $\delta$ Scuti stars than OP or OPLIB tables.

The mixing-length parameter $\alpha_{\mathrm{MLT}}$ is fixed at 1.89, following \cite{2012AJ....144...92Y}. Alternative mixing-length values have been proposed, for example, \cite{2023ApJ...942L..38D} suggest $\alpha_{\mathrm{MLT}}\approx0.5$ for Population I HADS stars. The sensitivity of model positions on the H–R diagram to variations in $\alpha_{\mathrm{MLT}}$ is tested and found minor shifts, consistent with \cite{1999MNRAS.302..349B}. The convective overshooting parameter is set as $f_\mathrm{ov}=0.015$ following \cite{2017MNRAS.467.3122N}, a choice that has a small effect on the models and supports the conclusion of \cite{2023MNRAS.526.1951D} that convective-core overshooting appears inefficient for evolved HADS stars. Rotational effects are also assessed and found to be negligible as $\rm{EH~Lib}$ is a slow rotator. Accordingly, the rotation velocity is fixed at 15 km s$^{-1}$.

Each evolutionary track commences from the zero-age main sequence and progresses through the post-MS stages, with the specified initial mass and chemical composition. At each evolutionary step, the pulsation frequencies are computed using the oscillation code GYRE.

\subsection{Parameter Fitting}\label{section:4.3}
To constrain the stellar models, the frequency of the fundamental mode $f_0$ detected from TESS light curves is used. Additionally, the period change rate $(1/P_0)(dP_0/dt)$ determined in this study falls within the range predicted for $\delta$ Scuti stars by \cite{1998A&A...332..958B}. Therefore, this rate can be attributed to stellar evolutionary effects, making it another valuable criterion for constraining models.

Figure~\ref{fig:Figure7} presents evolutionary tracks computed for the three chemical compositions from the zero-age main sequence to the end of post-MS stages. Models are calculated for masses in the range 1.50-2.50 $M_{\odot}$ with a mass step of 0.01 $M_{\odot}$. Following central hydrogen depletion ($X_c \leq 10^{-3}$), each evolutionary sequence is sampled at intervals of $10^5~\rm{yr}$. At every sampled model, theoretical frequencies of the eigenmodes are calculated with GYRE and the local period change rate is estimated from the frequency differences between adjacent time steps. Because the observed frequency uncertainty ($\approx 3\times10^{-6}~\rm{d^{-1}}$), is much smaller than the calculated uncertainty ($\approx 0.003~\rm{d^{-1}}$, here taken as the difference between consecutive time-step calculated frequencies), model points whose calculated fundamental frequency $f_\mathrm{0,cal}$ falls within the model uncertainty of the observed $f_0$ are marked in black. Model points for which the calculated period change rate $(1/P_0)(dP_0/dt)_\mathrm{cal}$ agrees with the observed value within $1\sigma$ are indicated in blue. Points satisfying both the frequency and period change rate criteria simultaneously are highlighted in red.

\begin{figure*}
    \centering
    \begin{subfigure}[H]{\textwidth}
        \centering
        \includegraphics[width=\textwidth]{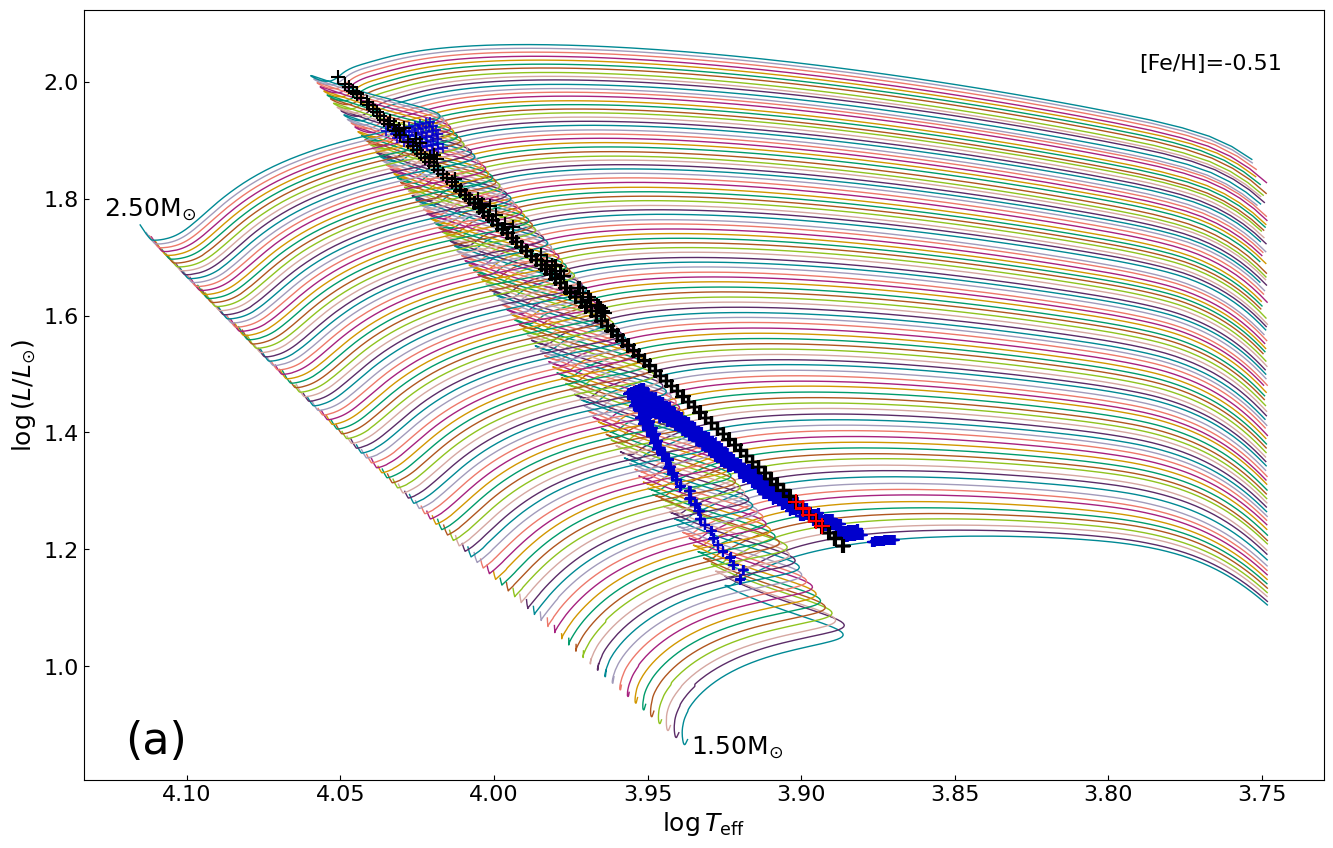}
        \label{fig:Figure7_a}
    \end{subfigure}

    \centering
    \begin{subfigure}[H]{\textwidth}
        \centering
        \includegraphics[width=\textwidth]{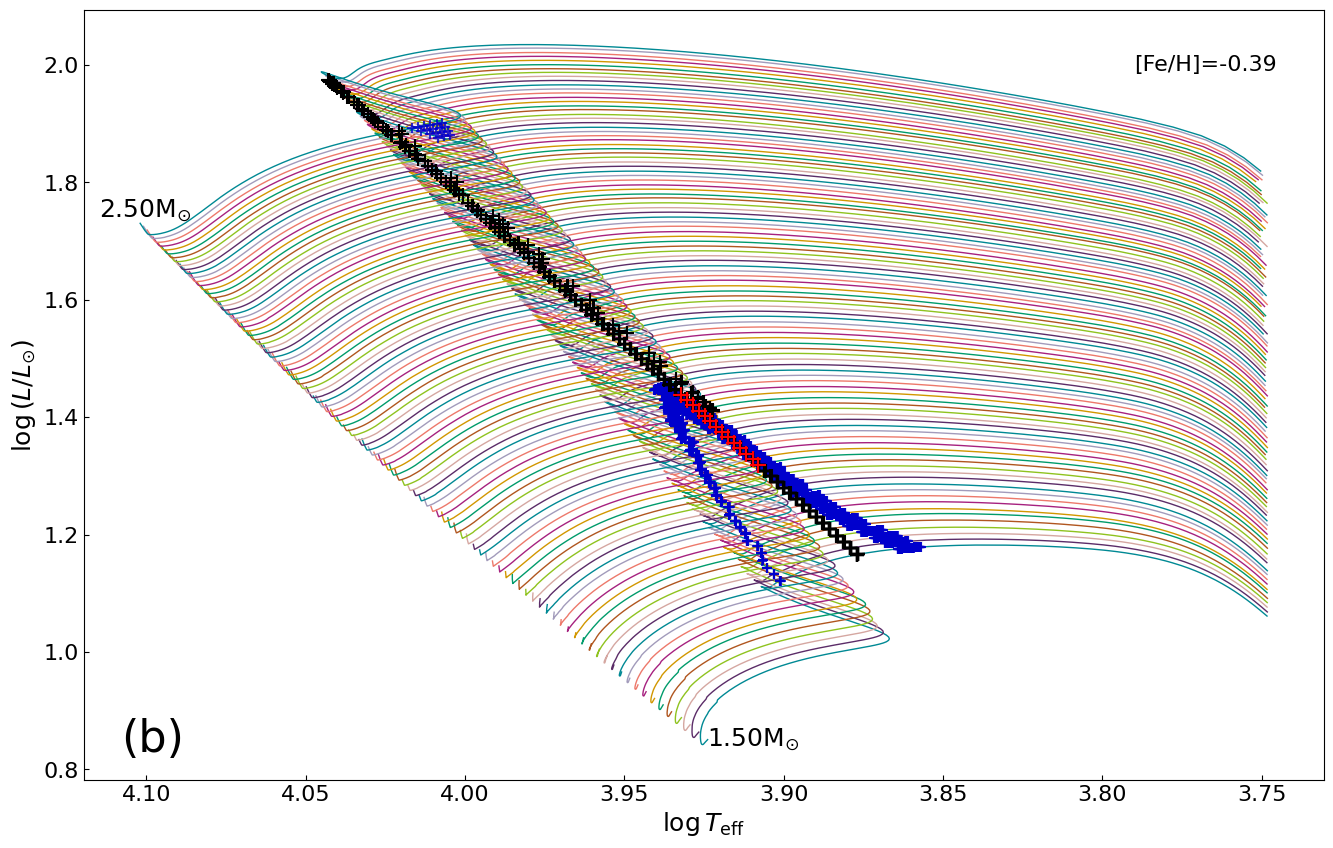}
        \label{fig:Figure7_b}
    \end{subfigure}
\end{figure*}

\begin{figure*}
    \ContinuedFloat
    \centering
    \begin{subfigure}[H]{\textwidth}
        \centering
        \includegraphics[width=\textwidth]{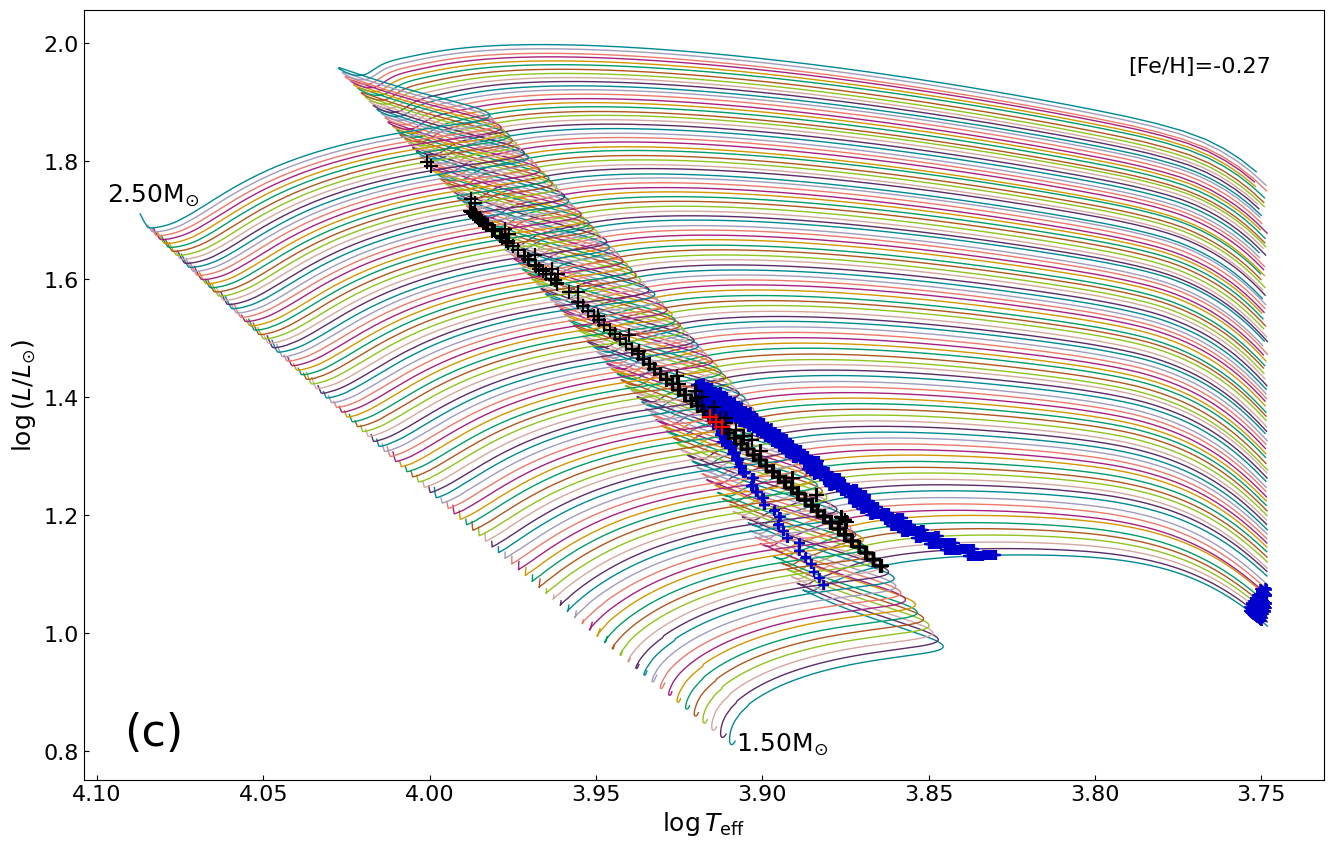}
        \label{fig:Figure7_c}
    \end{subfigure}
    \caption{Evolutionary tracks from the zero-age main sequence to the end of post-MS stages for three initial chemical compositions: (a) $(X,Y,Z)=(0.7384,0.2569,0.0047)$; (b) $(X,Y,Z)=(0.7343,0.2595,0.0062)$; (c) $(X,Y,Z)=(0.7290,0.2629,0.0081)$. Models are computed for masses from 1.50 $M_{\odot}$ to 2.50 $M_{\odot}$ with a mass step of 0.01 $M_{\odot}$. Following central hydrogen depletion, each evolutionary sequence is sampled at intervals of $10^5~\rm{yr}$, and theoretical frequencies of the eigenmodes with period change rates are calculated at each step. Points for which the calculated fundamental frequency $f_\mathrm{0,cal}$ lies within the adopted model frequency uncertainty of the observed $f_0$ are plotted in black. Points for which the calculated period change rate $(1/P_0)(dP_0/dt)_\mathrm{cal}$ agrees with the observed rate within $1\sigma$ are plotted in blue. Points satisfying both criteria simultaneously are highlighted in red.}
    \label{fig:Figure7}
\end{figure*}

In addition to the Gaia value $\mathrm{log}\,(L_\mathrm{Gaia}/L_{\odot})=1.11\pm0.03$, bolometric luminosities are independently determined as $\mathrm{log}\,(L/L_{\odot})=1.13$ and 1.17 respectively using the Gaia parallax $\varpi$, the \textit{V}-band absolute magnitude $M_v$, and bolometric corrections from the ATLAS9 and ATLAS12 spectral libraries \citep{2003IAUS..210P.A20C,2014dapb.book...39K} with the YBC tool \citep{2019A&A...632A.105C}. Of the three model sequences considered, only the sequence with initial composition $(X,Y,Z)=(0.7384,0.2569,0.0047)$ yields theoretical luminosities $(\mathrm{log}\,(L_\mathrm{theo}/L_{\odot}) \approx 1.26)$ consistent with these observed values, while the other two sequences produce much higher theoretical luminosities $(\approx1.37)$. Accordingly, $\mathrm{[Fe/H]}=-0.51$ is adopted for EH Lib, and subsequent analysis is restricted to models with this composition.

\begin{figure}
    \centering
    \includegraphics[width=\columnwidth]{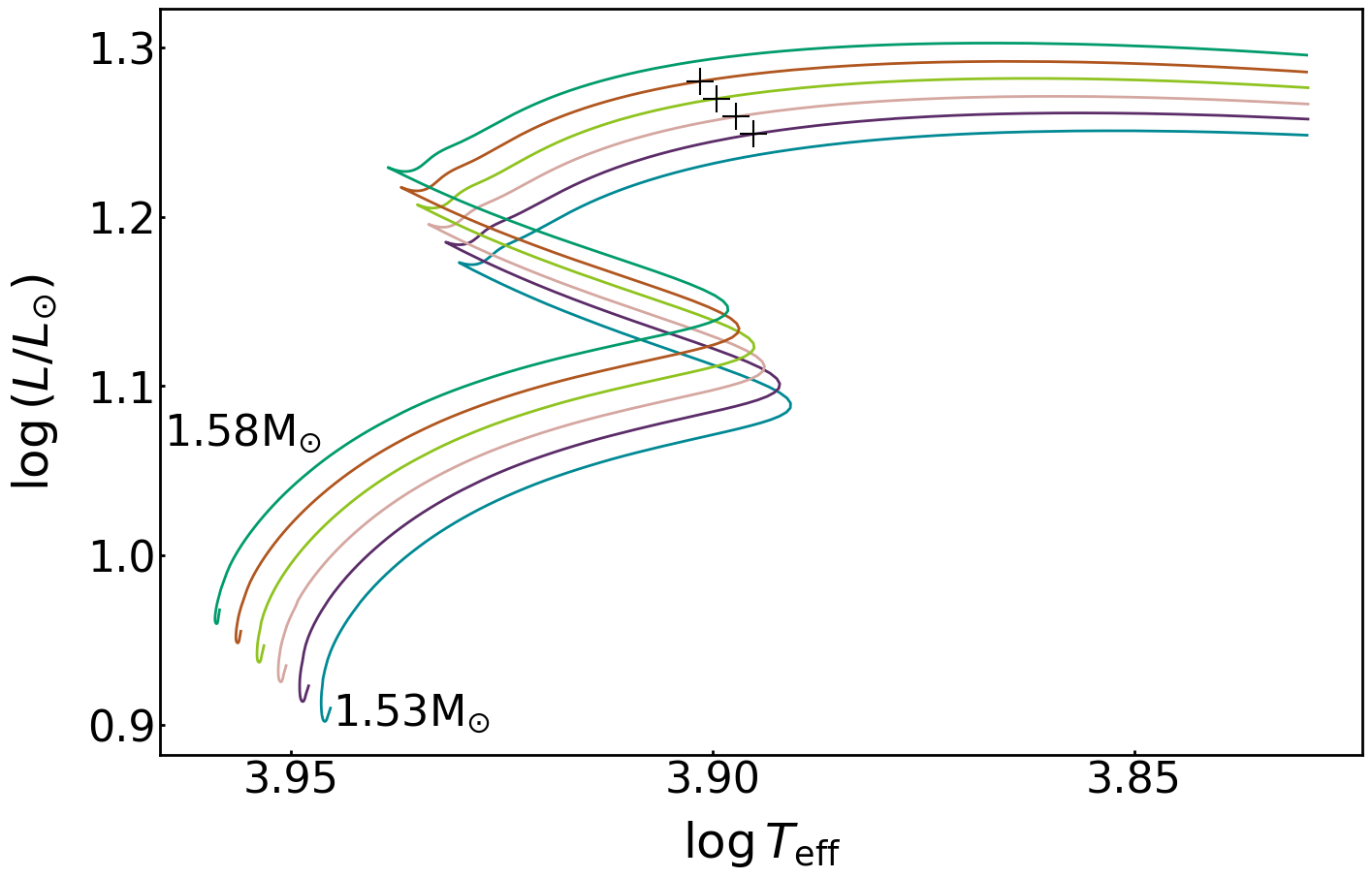}
    \caption{Evolutionary tracks from the zero-age main sequence to the end of post-MS stages in the mass range of 1.53-1.58 $M_{\odot}$. The black crosses mark the fitted models of $\rm{EH~Lib}$.}
    \label{fig:Figure8}
\end{figure}

The Details of the fitted models are provided in Table~\ref{tab:Tabel6}, and the derived stellar parameters of $\rm{EH~Lib}$ are summarized in Table~\ref{tab:Tabel7}. Figure~\ref{fig:Figure8} shows the positions of the fitted models on the HR diagram.

\begin{table*}
\centering
\renewcommand{\arraystretch}{1.2}
\caption{The fitted stellar models for $\rm{EH~Lib}$.}
\label{tab:Tabel6}
\begin{tabular}{cccccccc}
\hline
$M~(M_{\odot})$ & Age$~(10^{9}~\rm{yr})$ & $T_{\rm eff}~\rm{(K)}$ & $\mathrm{log}\,(L/L_{\odot})$ & log$~g~\rm{(dex)}$ & $f_0~\rm{(d^{-1})}$ & $\frac{1}{P_0}\frac{dP_0}{dt}~\rm{(\times10^{-9}~yr^{-1})}$ \\
\hline
1.54 & 1.507 & 7854 & 1.2487 & 3.9121 & 11.3127 & 5.06 \\
1.55 & 1.476 & 7892 & 1.2590 & 3.9129 & 11.3098 & 5.28 \\
1.56 & 1.449 & 7933 & 1.2697 & 3.9139 & 11.3133 & 5.57 \\
1.57 & 1.419 & 7971 & 1.2799 & 3.9147 & 11.3109 & 5.76 \\
\hline
\end{tabular}
\end{table*}

\begin{table}
\centering
\renewcommand{\arraystretch}{1.2}
\caption{The determined stellar parameters of $\rm{EH~Lib}$.}
\label{tab:Tabel7}
\begin{tabular}{p{90pt} p{60pt}}
\hline
Parameter & Value\\
\hline
$\rm{M~(M_{\odot}})$ & $1.555\pm0.015$ \\
Age$~(10^{9}~\rm{yr})$ & $1.46\pm0.05$\\
$T_{\rm eff}~(\rm{K})$ & $7913\pm58$\\
$\mathrm{log~g}~(\rm{dex})$ & $3.913\pm0.001$\\
$\mathrm{log}\,(L/L_{\odot})$ & $1.26\pm0.02$\\
\hline
\end{tabular}
\end{table}

\begin{figure}
    \centering
    \includegraphics[width=\columnwidth]{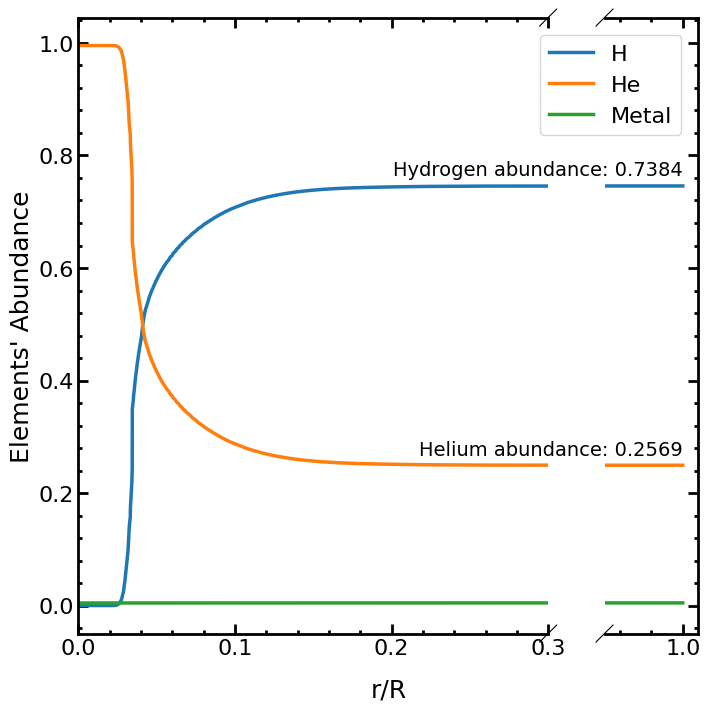}
    \caption{Internal distributions of hydrogen, helium, and metals plotted against fractional stellar radius for the 1.555 $M_{\odot}$ model.}
    \label{fig:Figure9}
\end{figure}

\begin{figure}
    \centering
    \includegraphics[width=\columnwidth]{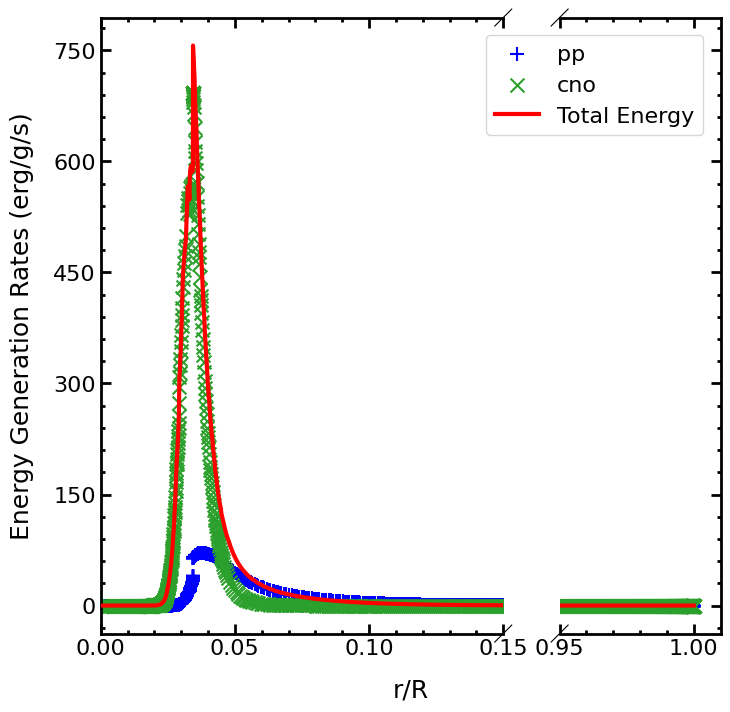}
    \caption{Internal distributions of energy generation rates plotted against fractional stellar radius for the 1.555 $M_{\odot}$ model.}
    \label{fig:Figure10}
\end{figure}

The internal structure and evolutionary stage of $\rm{EH~Lib}$ are investigated using these fitted stellar models. Taking the model with a mass of 1.555 $M_{\odot}$ as an example, Figures~\ref{fig:Figure9} and \ref{fig:Figure10} respectively illustrate the internal distributions of elemental abundances and energy generation rates against fractional stellar radius. These figures indicate that $\rm{EH~Lib}$ has exhausted hydrogen in its core, resulting in the formation of a helium core, surrounded by a hydrogen-burning shell where the energy is currently generated. This suggests that the star has evolved off the main sequence and is currently ascending the subgiant branch in the HR diagram.

\section{Discussion}\label{section:5}
Regarding the potential binarity of EH~Lib, \cite{1981AcASn..22..279J} obtained 6 new times of maximum light with a high-speed photoelectric photometer and, combining these with earlier data, first proposed the hypothesis that EH~Lib may reside in a binary system. Later, \cite{2019A&A...623A..72K} suggested proper motion variations that might hint at a possible companion, though they cautioned that the signal-to-noise ratio was too low for a firm conclusion. 

In this study, we investigated this hypothesis using two independent methods. First, our updated O-C analysis reveals no clear evidence of orbital motion. Second, we analyzed historical radial velocity (RV) data \citep{1986PASP...98..651J,1993PASP..105..809W}, excluding 11 data points from \cite{1976PASP...88..164M} due to the limited precision of their low-dispersion spectra. As shown in Figure~\ref{fig:FigureA3}, the RV variations are prominently dominated by the fundamental radial pulsation. After subtracting a 3-term Fourier fit of the pulsation model, the RV residuals plotted over BJD (Figure~\ref{fig:FigureA4}) scatter around zero with a standard deviation of 4.1~km/s and show no signal indicating binarity.

Based on the results of both the O-C and RV diagrams, we propose that there is no obvious companion near EH~Lib that could significantly influence the orbital motion. However, due to the limited time spans and observing gaps in both datasets, the existence of a long-period or low-mass companion cannot be entirely ruled out. To settle this issue definitively, further continuous, precise, and long-duration observations are required. If EH~Lib is eventually confirmed to be in a binary system, the independent orbital parameters would provide valuable additional constraints for the theoretical modeling.

Constraining stellar models for a single-radial-mode HADS star like $\rm{EH~Lib}$ meets significant challenges, as discussed in Section~\ref{section:3}, due to a large number of models match the observed frequencies within the uncertainties, as shown in Figure~\ref{fig:Figure7}. However, incorporating the period change rate and luminosity as additional constraints has proven effective, as demonstrated in Figures~\ref{fig:Figure7} and \ref{fig:Figure8}. These results also confirm that the observed period change rate of $\rm{EH~Lib}$ can be attributed to the stellar evolutionary effects.

Based on our best-fitting models with $\mathrm{[Fe/H]}=-0.51$, we further compared the theoretically constrained fundamental parameters with the observations. The surface gravity derived from the models, $\mathrm{log~g_\mathrm{theo}}=3.913\pm0.001~\rm{dex}$, is in excellent agreement with the observed value, $\rm{log~g_\mathrm{obs}}=3.9\pm0.1~\rm{dex}$. However, the theoretical luminosity, $\mathrm{log}\,(L_\mathrm{theo}/L_{\odot})=1.26\pm0.02$, is slightly higher than the values derived from Gaia parallax and photometry ($\mathrm{log}\,(L/L_{\odot})=1.11 \sim 1.17$ depending on different bolometric corrections). Similarly, the theoretical effective temperature, $T_{\rm eff,theo}=7913\pm58~\rm{K}$, is higher than the spectroscopic value, $T_{\rm eff,obs}=7300\pm100~\rm{K}$, reported by \cite{2017MNRAS.470.4408K}.

However, a detailed evaluation reveals that these apparent differences can be well understood by considering the intrinsic stellar variations and observational uncertainties. Regarding the effective temperature, as a HADS star, EH~Lib experiences dramatic thermodynamic variations during its pulsation cycle. Following the Stefan-Boltzmann law, its large luminosity amplitude is accompanied by significant effective temperature fluctuations. We calculate that the theoretical $T_{\rm eff}$ variation over a full pulsation cycle for our best-fitting model is approximately $1000~\rm{K}$. Because the spectroscopic observation in the literature lacks a specific pulsation phase timestamp, the single-epoch spectrum could have been captured during a cooler phase, naturally explaining its deviation from our theoretical mean state.

Furthermore, the formal observational errors likely underestimate the true uncertainties. The quoted $\pm 100~\rm{K}$ error for the observed $T_{\rm eff}$ merely reflects the step size of the spectrum synthesis method, whereas realistic uncertainties are generally larger due to factors such as inaccurate atomic data, non-local thermodynamic equilibrium effects, and continuum normalization inaccuracies \citep{2013MNRAS.431.3685T}. For the luminosity, the derivation is susceptible to systematic errors. It strongly depends on the adopted extinction coefficients and the bolometric correction. Because bolometric correction is sensitive to $T_{\rm eff}$, any underestimated uncertainty in temperature propagates into the observed luminosity calculation. The derivation also relies on the choice of empirical models, as demonstrated in Section~\ref{section:4.3}, where adopting the exact same $T_{\rm eff}$ with different models yielded distinctly different $L$ estimates.

When the intrinsic temperature variation and the systematic uncertainties in deriving $L$ and $T_{\rm eff}$ are properly accounted for, the theoretical and observed parameters are, in fact, consistent within realistic physical bounds. This consistency further confirms the reliability of our purely asteroseismic analysis as an independent and accurate method for deriving fundamental stellar parameters.

For the frequencies $f_1$ and $f_2$ listed in Table~\ref{tab:Tabel2}, the observed values are $f_\mathrm{1,obs}=21.107\pm 0.001~\rm{d^{-1}}$ and $f_\mathrm{2,obs}=10.750\pm 0.002~\rm{d^{-1}}$. Supported by our independent ground-based photometry, we confirm that these two frequencies are not due to contamination from nearby fainter sources. Our analysis indicates that $\rm{EH~Lib}$ is an evolved star, whose internal structure allows for the existence of mixed modes. Mixed modes are single pulsation modes that exhibit a dual character, which behave like acoustic waves (p-modes) in the stellar envelope and like gravity waves (g-modes) in the core. This phenomenon arises from the coupling between p- and g-mode propagation cavities within evolved stars \citep{1977A&A....58...41A}. Consequently, a single mixed mode is best described not by a single radial order $n$, but by the acoustic-wave winding number $n_\mathrm{p}$ and the gravity-wave winding number $n_\mathrm{g}$. These numbers serve as the effective radial orders in the respective propagation cavities.

For $f_1$, our fitted models consistently identify it as a mixed mode. Table~\ref{tab:Tabels3} lists the theoretical frequencies from our fitted models listed in Table~\ref{tab:Tabel6} that are closest to $f_\mathrm{1,obs}$. The harmonic degree is restricted to $l<4$ due to the disc-averaging effect \citep{1977AcA....27..203D}, and the azimuthal order is set to $m=0$ as rotational effects are negligible. All models identify $f_1$ as a mode with a clear p-mode character of $n_\mathrm{p}=3$. The high gravity-wave winding number confirms its mixed-mode nature and provides strong evidence that $\rm{EH~Lib}$ is an evolved star. The frequency $f_2$ satisfies $f_2 \approx f_1-f_0+1~\rm{d^{-1}}$. Given its much lower amplitude compared to $f_0$, and $\rm{EH~Lib}$’s proximity to the ecliptic, $f_2$ may be attributed to the combination of $f_0$, $f_1$ and the instrumental effects, such as scattered light contamination or momentum dumps. Similar features have been observed in other HADS stars during the same TESS sector such as $\rm{TIC~16283570}$, providing cautious supports to this interpretation.

\begin{table}
\centering
\renewcommand{\arraystretch}{1.2}
\caption{The calculated frequencies $f_\mathrm{1,cal}$ of the fitted models listed in Table~\ref{tab:Tabel6} that are closest to the observed frequency $f_\mathrm{1,obs}$. Here, $n_\mathrm{p}$ denotes the acoustic-wave winding number, $n_\mathrm{g}$ the gravity-wave winding number, and $l$ the harmonic degree. $\Delta f=f_\mathrm{1,cal}-f_\mathrm{1,obs}$.}
\label{tab:Tabels3}
\begin{tabular}{cccccc}
\hline
$M~(M_{\odot})$ & $f_\mathrm{1,cal}~\rm{(d^{-1})}$ & $n_\mathrm{p}$ & $n_\mathrm{g}$ & $l$ & $\Delta$$f~\rm{(d^{-1})}$\\
\hline
1.54 & 21.144 & 3 & 11 & 1 & 0.037  \\
1.55 & 21.167 & 3 & 19 & 2 & 0.060  \\
1.56 & 21.106 & 3 & 19 & 2 & -0.001  \\
1.57 & 21.101 & 3 & 19 & 2 & -0.006  \\
\hline
\end{tabular}
\end{table}

\begin{figure*}
    \centering
    \includegraphics[width=\textwidth]{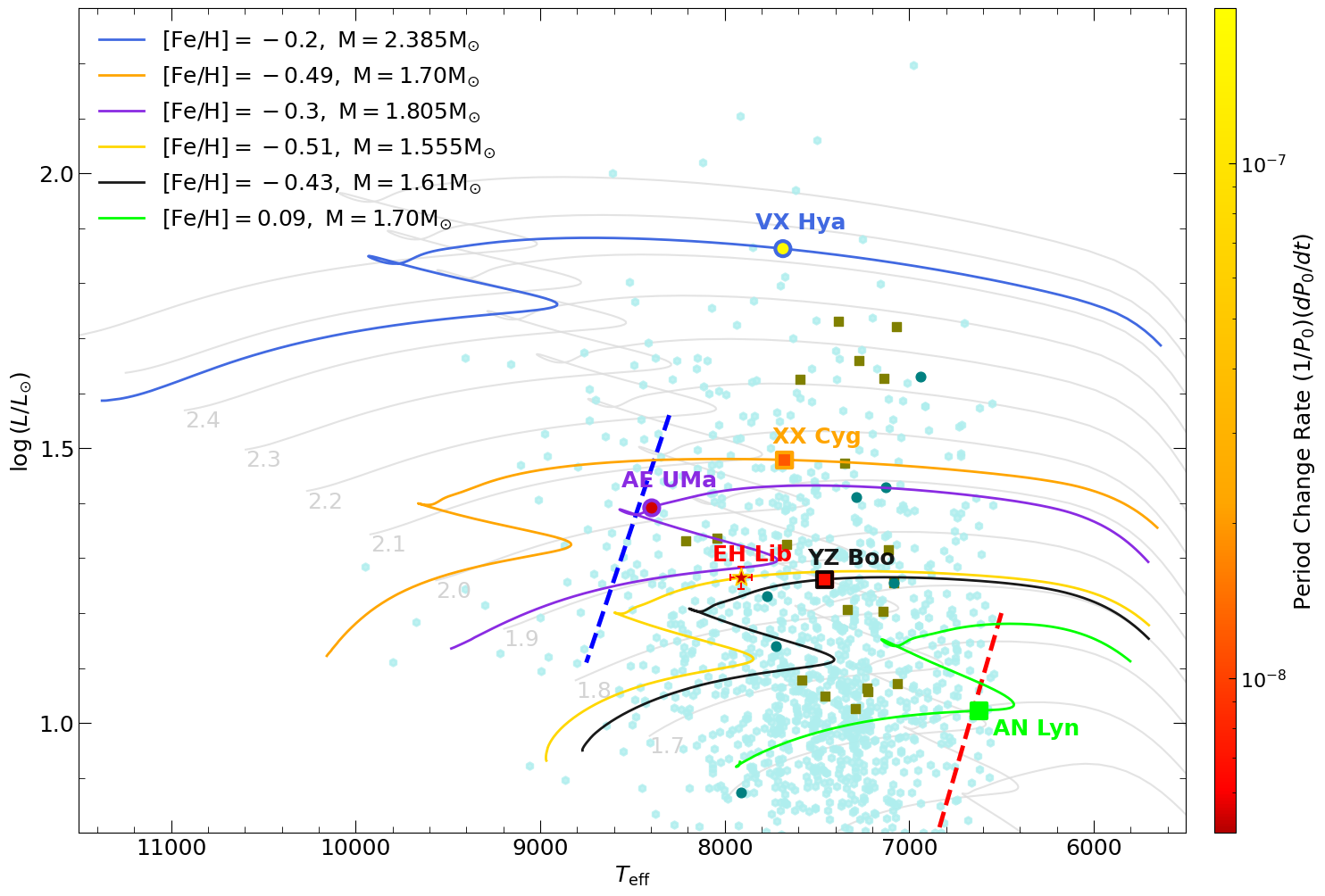}
    \caption{HR diagram for $\delta$ Scuti stars. The best-fit models of six HADS stars studied through asteroseismology are shown with distinct stroke colours and symbols, along with their corresponding evolutionary tracks. Fill colours indicate the period change rates, with redder colour corresponding to shorter rates. See Table~\ref{tab:Tabels1} for details. Light cyan hexagons present the sample of normal $\delta$ Scuti stars from \protect\cite{2019MNRAS.485.2380M}, selected based on fractional parallax uncertainties below 3 per cent and pulsation full amplitudes exceeding 0.01 mag. Squares represent single-radial-mode HADS stars, whereas circles represent multi-radial-mode HADS stars. The effective temperatures of the olive and dark cyan data points are derived from the spectral surveys LAMOST \protect\citep{2012RAA....12..723Z,2015MNRAS.448..822X} and GALAH \protect\citep{2015MNRAS.449.2604D,2021MNRAS.506..150B}. The luminosities are provided by Gaia DR3 \protect\citep{2023A&A...674A...1G} and reduced following the method described in Section~\ref{section:3.2}. The HADS star sample is taken from an in-house program. To place these stars meaningfully, evolutionary tracks in the mass range 1.4-2.6 $M_{\odot}$ for the `standard model' \protect\citep{2019MNRAS.485.2380M} with the chemical composition ($X=0.71,~Z=0.014$), are computed with MESA. These tracks are shown as grey lines, with labelled masses at the beginnings. The blue and red dash lines mark the boundaries of the theoretical instability strip, derived using time-dependent convection and $\alpha_{\mathrm{MLT}}=1.8$ from \protect\cite{2005A&A...435..927D}.}
    \label{fig:Figure11}
\end{figure*}

Furthermore, the analysis reveals that $\rm{EH~Lib}$ is a normal HADS star pulsating in its fundamental mode and locating at the post-MS stage. This finding is consistent with the tendency proposed by \cite{2018ApJ...861...96X}, who suggests that the HADS stars which have left the main sequence and become subgiants with the hydrogen-burning shell, tend to exhibit lower fundamental frequencies as they evolve. 

Figure~\ref{fig:Figure11} reproduces Figure 9 of \cite{2018ApJ...861...96X}, presenting an HR diagram for $\delta$ Scuti stars, with $\rm{EH~Lib}$ plotted as a star and error bars indicating the uncertainties in effective temperature and luminosity. Table~\ref{tab:Tabels1} lists the observed fundamental frequencies, the period change rates, and the physical parameters of the best-fit models of these six HADS stars studied through asteroseismology \citep{2012AJ....144...92Y,2017MNRAS.467.3122N,2018MNRAS.473..398L,2018ApJ...861...96X,2018RAA....18....2Y}. These six HADS stars are found to be slightly evolved, crossing the instability strip. In addition to the trend reported by \cite{2018ApJ...861...96X}, it is suggested that the period change rate increases as the stars evolve.

\begin{table*}
\centering
\renewcommand{\arraystretch}{1.2}
\caption{The observed frequencies of the fundamental mode, period change rates and physical parameters of the best-fit models of six HADS stars studied through asteroseismology.}
\label{tab:Tabels1}
\begin{tabular}{lrrllll}
\hline
Star & $f_0~(\rm{d^{-1}})$ & $(1/P_0)(dP_0/dt)\mathrm{(yr^{-1})}$ & $M/M_{\odot}$ & Age ($10^9$ years) & [Fe/H] & Reference \\
\hline
$\rm{VX~Hya}$ & 4.4763 & $1.81(\pm0.09)\times10^{-7}$ & 2.385 & 0.43 & -0.2 & \cite{2018ApJ...861...96X} \\
$\rm{XX~Cyg}$ & 7.4148 & $1.19(\pm0.13)\times10^{-8}$ & 1.70 & 0.9 & -0.49 &  \cite{2012AJ....144...92Y}\\
$\rm{YZ~Boo}$ & 9.6069 & $6.7(\pm0.9)\times10^{-9}$ & 1.61 & 1.44 & -0.43 & \cite{2018RAA....18....2Y} \\
$\rm{AN~Lyn}$ & 10.1721 &  & 1.70 & 1.33 & 0.09 & \cite{2018MNRAS.473..398L} \\
$\rm{EH~Lib}$ & 11.3105 & $5.4(\pm0.5)\times10^{-9}$ & 1.555 & 1.46 & -0.51 & this work \\
$\rm{AE~UMa}$ & 11.6256 & $5.4(\pm1.9)\times10^{-9}$ & 1.805 & 1.055 & -0.3 & \cite{2017MNRAS.467.3122N} \\
\hline
\end{tabular}
\end{table*}

Notably, with $\rm{EH~Lib}$ constrained to $M=1.555\pm0.015~M_{\odot}$, this study provides independent empirical evidence that the instability domain of HADS stars extends to at least $\sim 1.56~M_{\odot}$, thereby tightening observational constraints on the low-mass boundary. Consequently, stellar evolution and oscillation models, and the physical inputs employed, should be capable of reproducing both the mode excitation and selection observed at such low masses. To corroborate this conclusion, further high-precision spectroscopy to refine $T_{\rm eff}$ and [Fe/H] together with extended asteroseismic monitoring are necessary.

\section{Conclusions}\label{section:6}
Through the analysis of space-based photometry from the TESS mission, we extract 17 significant frequencies, comprising fundamental frequency $f_0$, newly detected frequencies $f_1$ and $f_2$, in addition to the harmonics of $f_0$ and a linear combination, and determine 158 times of maximum light. Ground-based photometric data collected in the \textit{V} and \textit{R} bands between 2009 and 2018 provide additional 91 times of maximum light. Based on the pulsation constant and P-L relations, $f_0$ is identified as the fundamental radial pulsation mode.

The classical O−C method is utilized to estimate the period change rate of $\rm{EH~Lib}$. The period change rate of the fundamental mode which is consistent with that of the star, is determined as $(1/P_0)(dP_0/dt)=(5.4\pm0.5)\times10^{-9}~\mathrm{yr^{-1}}$, based on 342 times of maximum light spanning over 70 years. Additionally, a new ephemeris is derived:

\begin{equation}
    C = \rm{BJD}~2435223.7596 + 0.088413248 \times E + 5.76 \times 10^{-14}~E^2
    \nonumber
\end{equation}

The stellar evolutionary models are constructed with initial masses ranging from 1.50 $M_{\odot}$ to 2.50 $M_{\odot}$ and the metal abundances $Z=0.0047,0.0062$ and $0.0081$. By constraining these models using the observed $f_0$, $(1/P_0)(dP_0/dt)$ and luminosity within respective uncertainties, we determine the stellar parameters of $\rm{EH~Lib}$ (see Table~\ref{tab:Tabel7}). The results indicate that $\rm{EH~Lib}$ is a single-radial-mode HADS star, locating at the subgiant branch, with a helium core and a hydrogen-burning shell.

Furthermore, we would like to emphasize that: (a) for $\rm{EH~Lib}$, the period change rate can be attributed to stellar evolutionary effects; (b) this study of $\rm{EH~Lib}$ confirms that constraining stellar models for HADS stars using both frequencies and period change rates is feasible; (c) the determined position of $\rm{EH~Lib}$ on the HR diagram indicates that it is in the post-MS stage, consistent with the general consensus for HADS stars; (d) this study, with $\rm{EH~Lib}$ constrained to $M=1.555\pm0.015~M_{\odot}$, shows the instability domain of HADS stars can extend to at least $\sim 1.56~M_{\odot}$, validating that current stellar evolution and pulsation models could reproduce high-amplitude radial excitation and mode selection at such low masses. Therefore, this work lays the groundwork for future studies of the commonalities and specific properties to deepen our understanding of HADS stars.

\begin{acknowledgments}
JNF acknowledges the support from the National Natural Science Foundation of China (NSFC) through the grant 12427804. This work is supported by the China Manned Space Program with grant no. CMS-CSST-2025-A13, the Tianchi Talent Introduction Plan, and the Central Guidance for Local Science and Technology Development Fund under No. ZYYD2025QY27. LFM. and JNF acknowledge financial support from the UNAM under grant PAPIIT IN117323. HFX acknowledges support from the National Natural Science Foundation of China (NSFC) (No. 12303036). JS acknowledges the support from the International Centre of Supernovae, Yunnan Key Laboratory (No. 202302AN360001) and the West Light Foundation of the Chinese Academy of Sciences. This work is based upon observations carried out at the Yunnan Astronomical Observatory of China, Xinglong station of National Astronomical Observatories of China, and Observatorio Astron\'omico Nacional on the Sierra San Pedro M\'artir (OAN-SPM), Baja California, M\'exico. We thank the technical staff and night assistants for facilitating and helping making the observations. This work makes use of data collected by the TESS mission, which are publicly available from the Mikulski Archive for Space Telescopes (MAST, \url{https://archive.stsci.edu}). All the TESS data presented in this article were obtained from the MAST at the Space Telescope Science Institute. The specific observations analyzed can be accessed via \dataset[10.17909/dfqc-6p17]{https://doi.org/10.17909/dfqc-6p17}. Funding for the TESS mission is provided by NASA’s Science Mission directorate. This work also includes data from the European Space Agency (ESA) mission Gaia, (\url{https://www.cosmos.esa.int/gaia}), processed by the Gaia Data Processing and Analysis Consortium (DPAC, \url{https://www.cosmos.esa.int/web/gaia/dpac/consortium}). Funding for the DPAC has been provided by national institutions, in particular the institutions participating in the Gaia Multilateral Agreement. This research made use of Lightkurve, a Python package for Kepler and TESS data analysis. We gratefully acknowledge them for making such high-quality data available. We also acknowledge the use of ChatGPT (OpenAI) and Gemini (Google) for text editing and language polishing of this manuscript. Finally, we are indebted to Dr. Filiz Kahraman Ali{\c{c}}avu{\c{s}} for insightful discussions on the spectroscopic data and to Dr. Timothy R. Bedding, Dr. Peter De Cat as well as Dr. Tianqi Cang for valuable suggestions on this manuscript. Our gratitude also extends to the anonymous reviewer for the constructive feedback during the peer-review process.
\end{acknowledgments}

%
\facilities{TESS, Gaia, MAST}

\software{IRAF \citep{1986SPIE..627..733T,1993ASPC...52..173T}, AstroImageJ \citep{2017AJ....153...77C}, Period04 \citep{2005CoAst.146...53L}, MESA \citep{2011ApJS..192....3P,2013ApJS..208....4P,2015ApJS..220...15P,2018ApJS..234...34P,2019ApJS..243...10P,2023ApJS..265...15J}, GYRE \citep{2013MNRAS.435.3406T,2018MNRAS.475..879T,2020ApJ...899..116G}}


\appendix

Figures~\ref{fig:FigureA1} and \ref{fig:FigureA2} display the whole resulting light curves after data reduction from the ground-based observations between 2009 and 2018 in \textit{V} and \textit{R} bands, respectively.

Figures~\ref{fig:FigureA3} and \ref{fig:FigureA4} show respectively the phase-folded radial velocity curve of EH~Lib based on literature data and the corresponding radial velocity residuals after subtracting the dominant pulsation.

Tables~\ref{tab:TabelA1},~\ref{tab:TabelA2} and~\ref{tab:TabelA3} present the times of maximum light for $\rm{EH~Lib}$ derived from new ground-based R-band observations, TESS observations, and previous literature, respectively.

\begin{figure*}
    \centering
    \begin{subfigure}[b]{\textwidth}
        \centering
        \includegraphics[width=\textwidth]{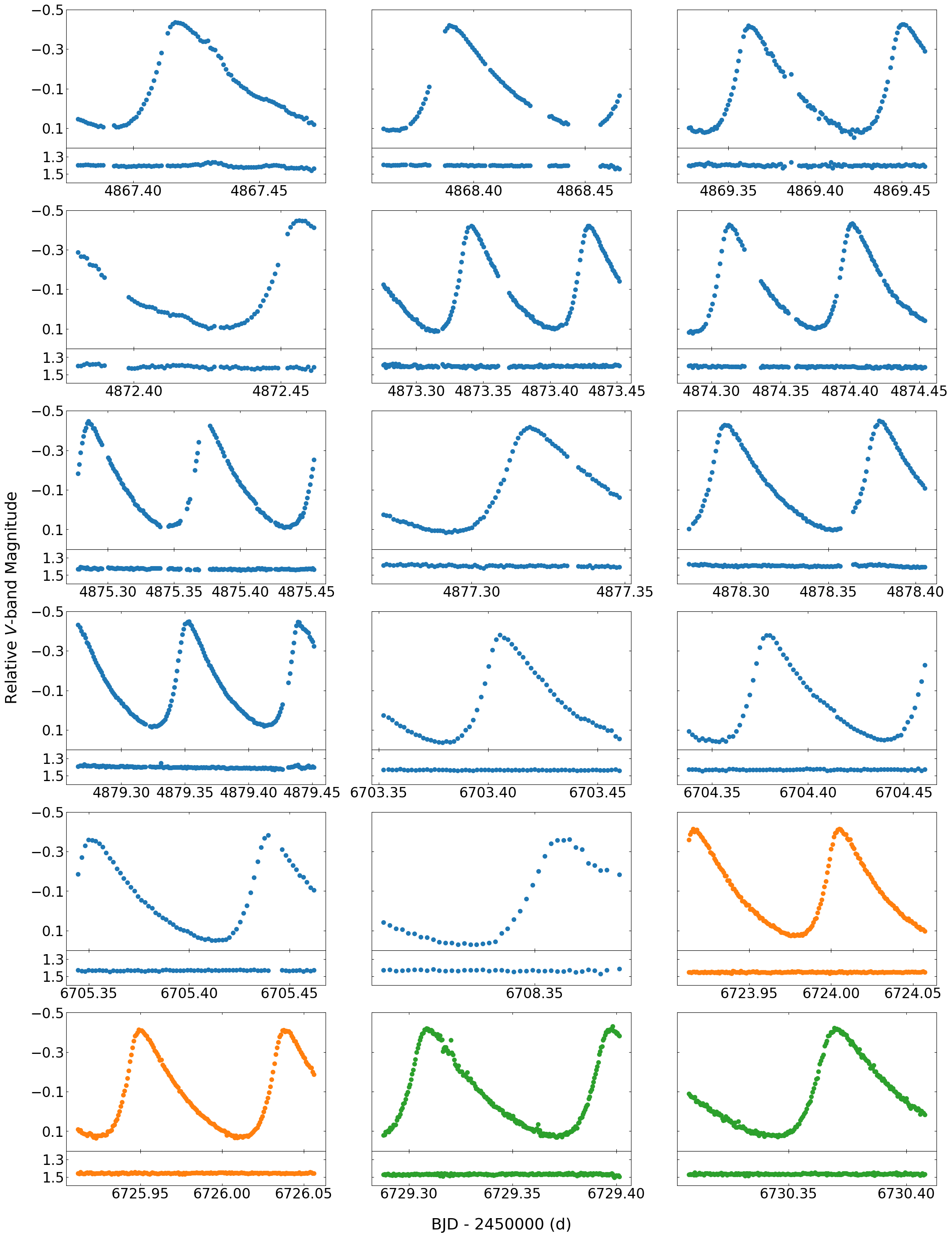}
        \label{fig:FigureA1_1}
    \end{subfigure}
\end{figure*}

\begin{figure*}
    \ContinuedFloat
    \centering
    \begin{subfigure}[b]{\textwidth}
        \centering
        \includegraphics[width=\textwidth]{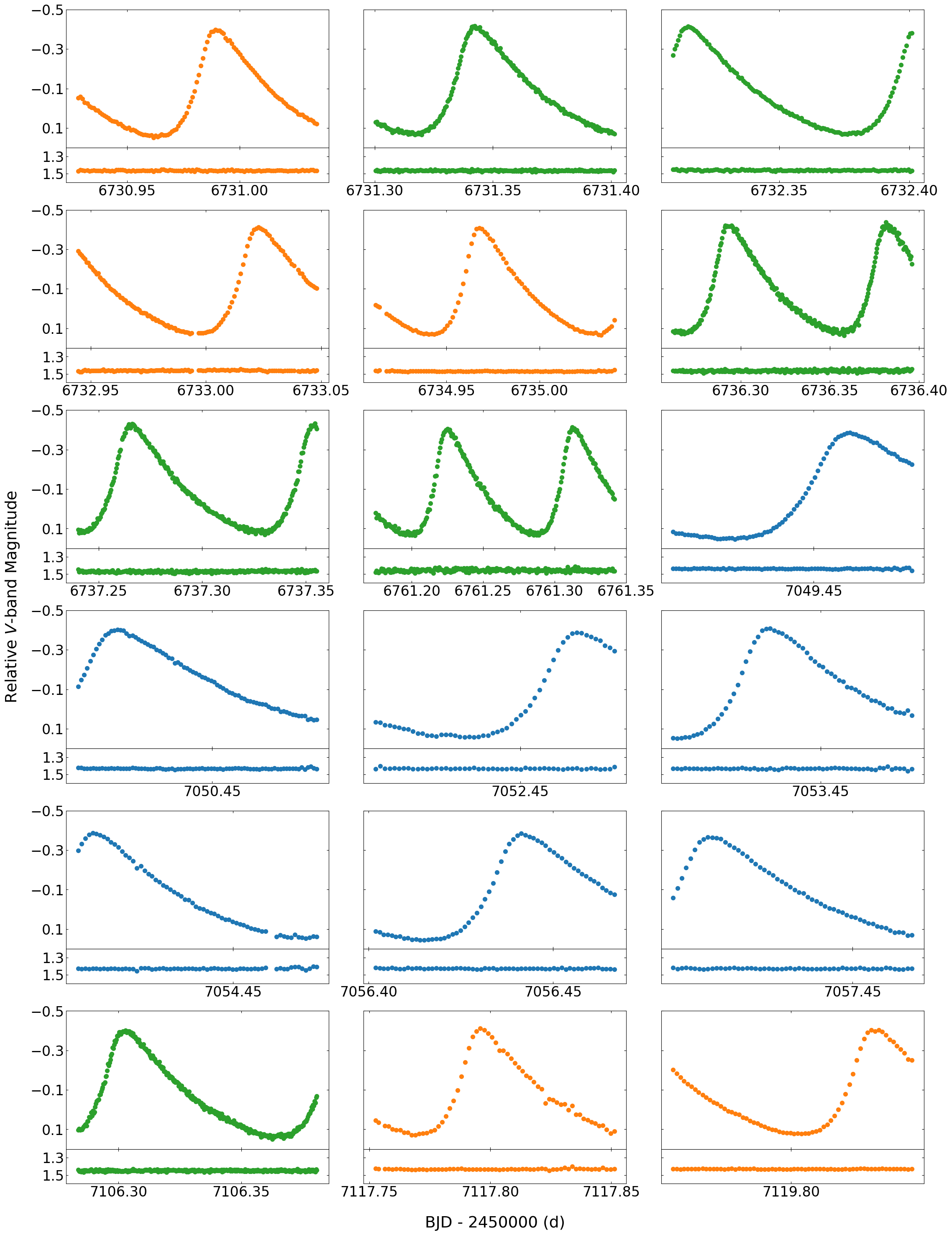}
        \label{fig:FigureA1_2}
    \end{subfigure}
\end{figure*}

\begin{figure*}
    \ContinuedFloat
    \centering
    \begin{subfigure}[b]{\textwidth}
        \centering
        \includegraphics[width=\textwidth]{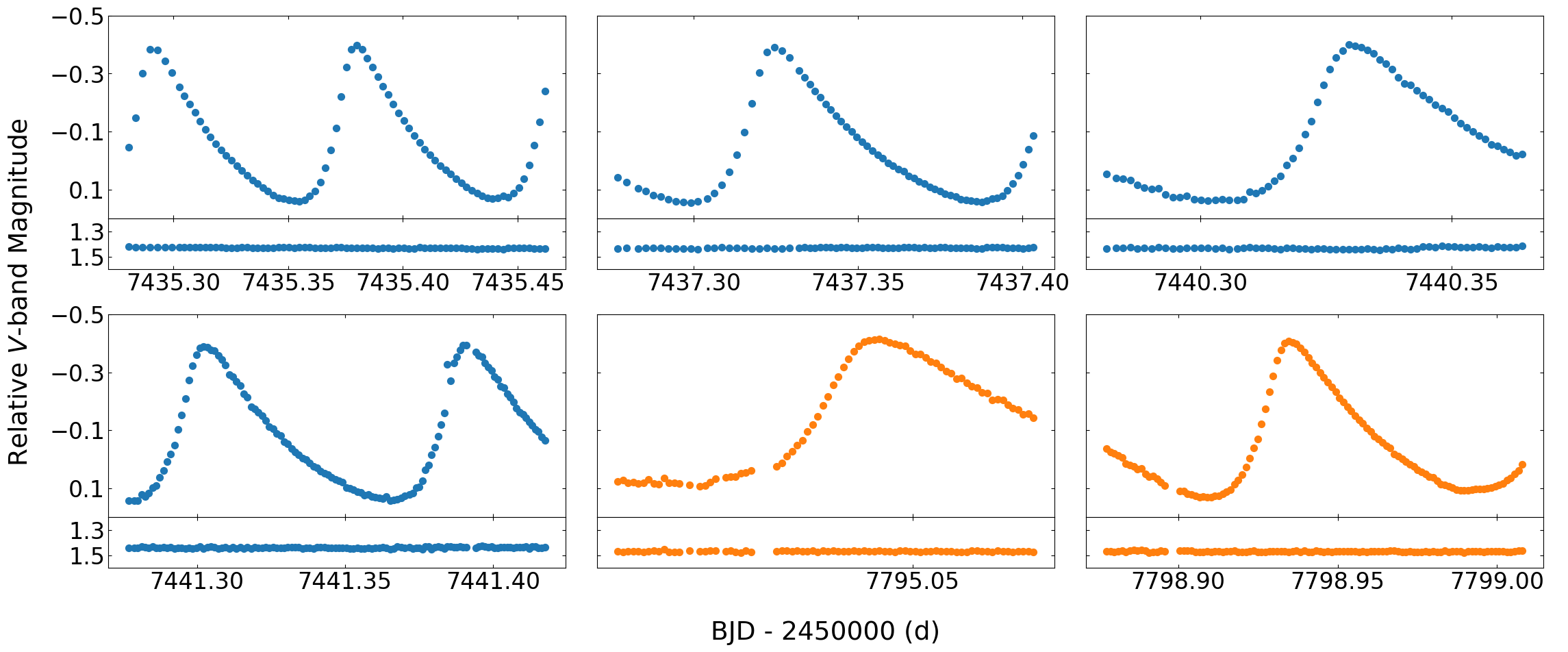}
        \label{fig:FigureA1_3}
    \end{subfigure}
    \caption{The light curves of $\rm{EH~Lib}$ in \textit{V}-band observed between 2009 and 2018. In each subfigure, the upper panel shows the magnitude differences between $\rm{EH~Lib}$ and the comparison star, and the bottom panel shows the magnitude differences between the check star and the comparison star. Outliers are removed. The blue points come from the observations with the 101.6-cm Telescope at Yunnan Astronomical Observatory of China, the orange ones from those with the 84-cm Telescope at National Astronomical Observatory San Pedro Mártir of Mexico, the green ones from those with the 85-cm Telescope at Xinglong Station, and the red ones from those with the Xinglong 2.16-m Telescope.}
    \label{fig:FigureA1}
\end{figure*}

\begin{figure*}
    \centering
    \begin{subfigure}[b]{\textwidth}
        \centering
        \includegraphics[width=\textwidth]{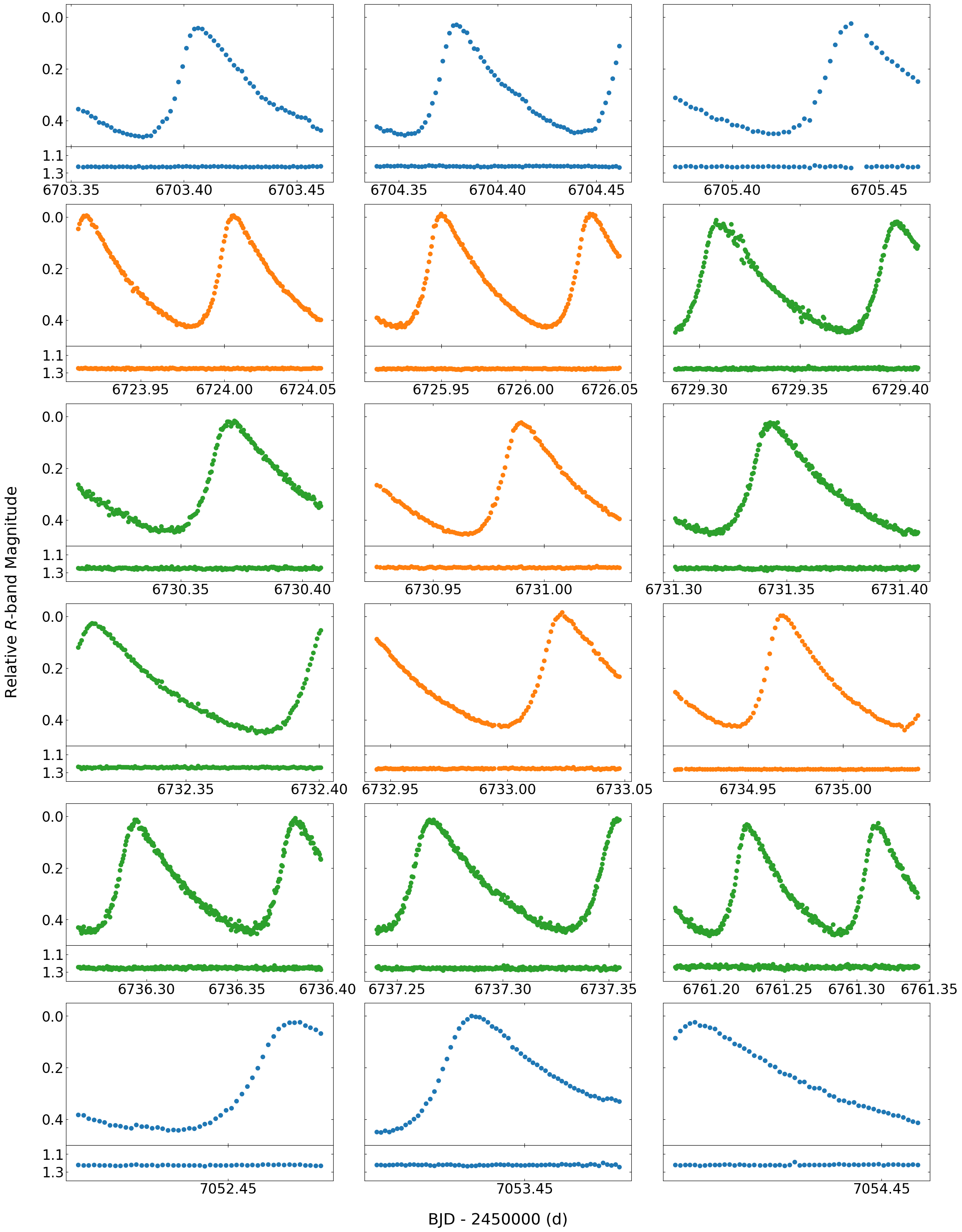}
        \label{fig:FigureA2_1}
    \end{subfigure}
\end{figure*}

\begin{figure*}
    \ContinuedFloat
    \centering
    \begin{subfigure}[b]{\textwidth}
        \includegraphics[width=\textwidth]{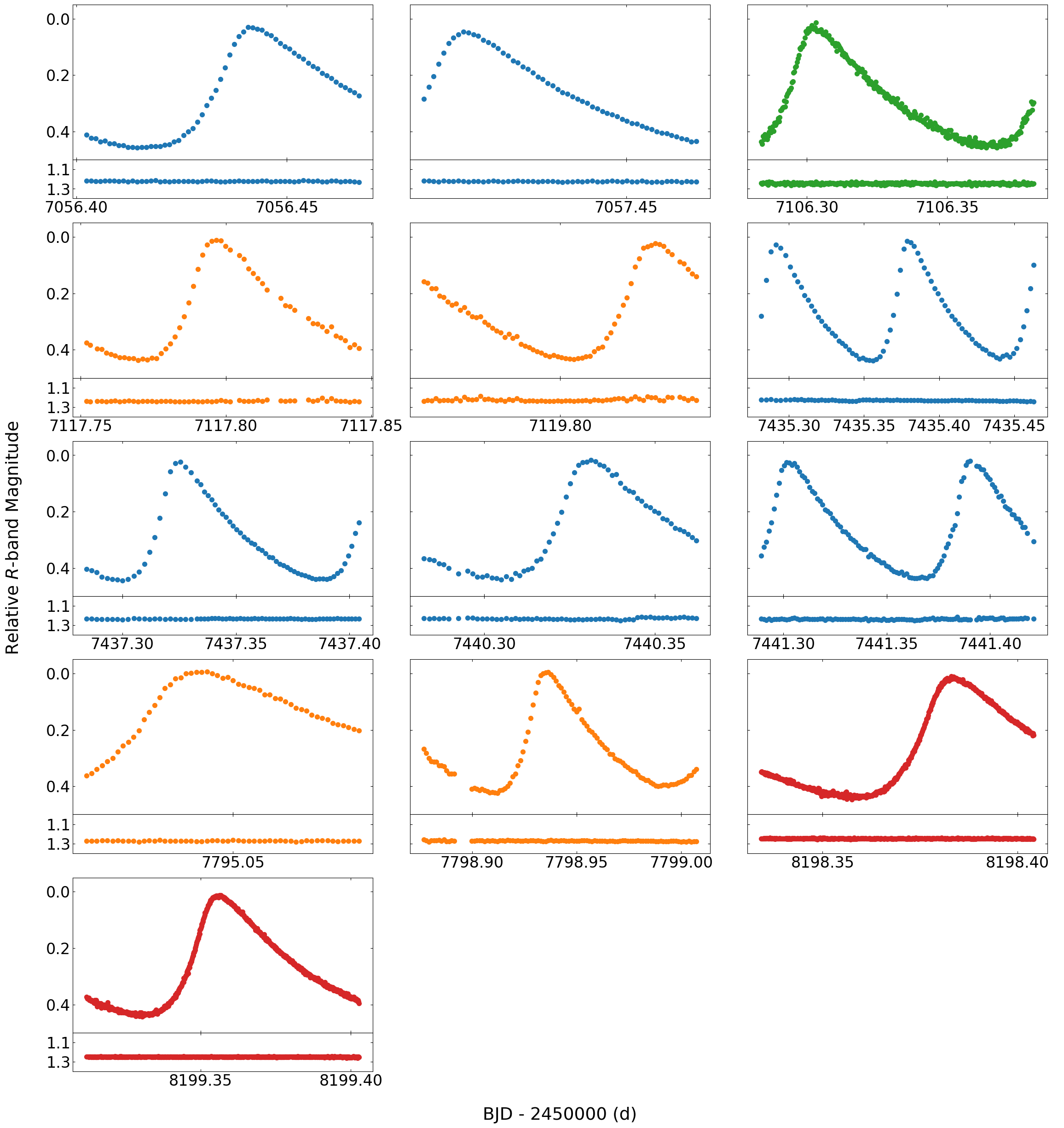}
        \label{fig:FigureA2_2}
    \end{subfigure}
    \caption{The light curves of $\rm{EH~Lib}$ in \textit{R}-band observed between 2009 and 2018. In each subfigure, the upper panel shows the magnitude differences between $\rm{EH~Lib}$ and the comparison star, and the bottom panel shows the magnitude differences between the check star and the comparison star. Outliers are removed. The blue points come from the observations with the 101.6-cm Telescope at Yunnan Astronomical Observatory of China, the orange ones from those with the 84-cm Telescope at National Astronomical Observatory San Pedro Mártir of Mexico, the green ones from those with the 85-cm Telescope at Xinglong Station, and the red ones from those with the Xinglong 2.16-m Telescope.}
    \label{fig:FigureA2}
\end{figure*}

\begin{figure}
\centering
\includegraphics[width=\textwidth]{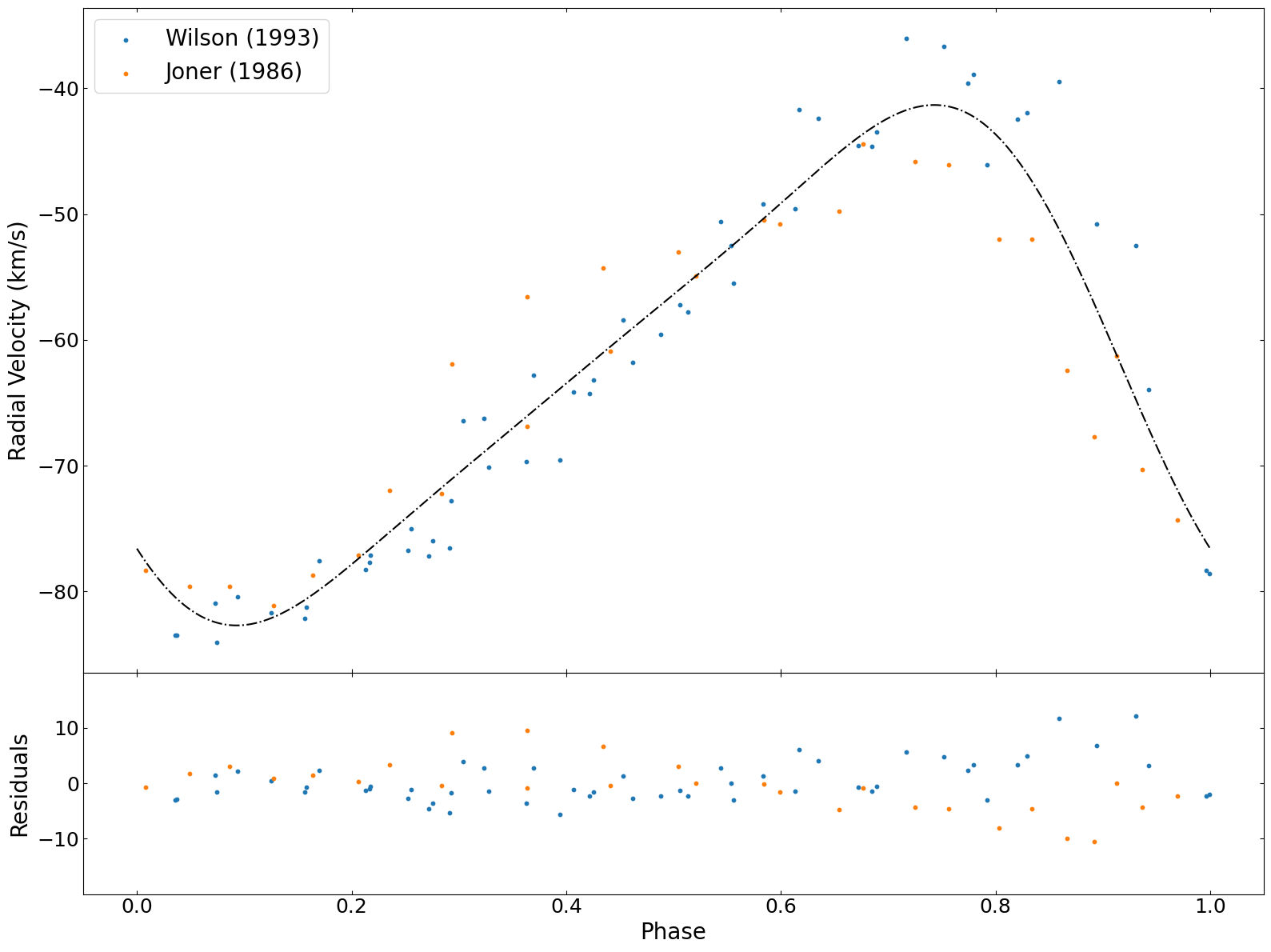}
\caption{Phase-folded radial velocity curve of EH~Lib using literature data (blue points from \cite{1993PASP..105..809W}; orange points from \cite{1986PASP...98..651J}). The black dashed line represents a 3-term Fourier fit to the dominant pulsation mode, and the bottom panel shows the corresponding residuals.}
\label{fig:FigureA3}
\end{figure}

\begin{figure}
\centering
\includegraphics[width=\textwidth]{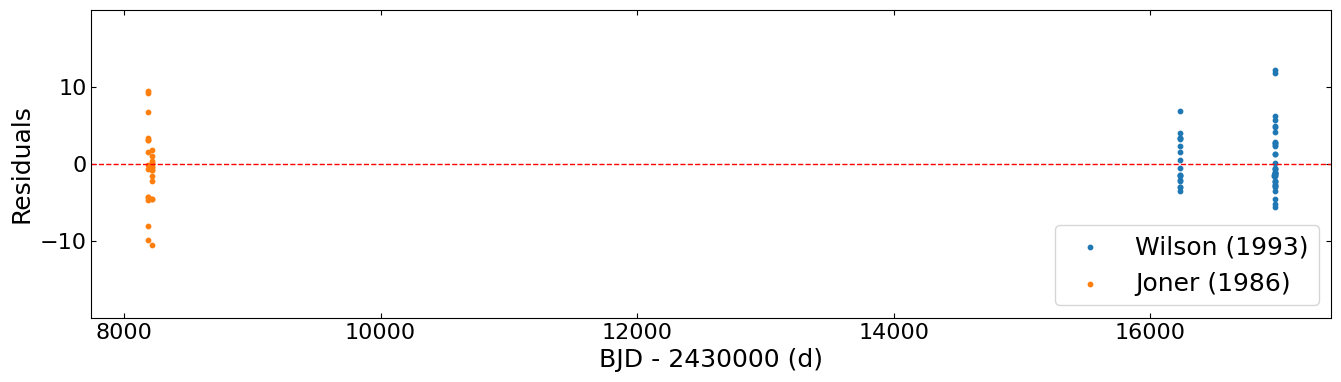}
\caption{Radial velocity residuals of EH~Lib (after subtracting the dominant pulsation model) plotted against BJD. The red dashed line denotes the zero mean. No obvious signal of orbital motion is observed.}
\label{fig:FigureA4}
\end{figure}

\begin{table*}
\centering
\renewcommand{\arraystretch}{1.2}
\setlength{\tabcolsep}{4pt}
\caption{Times of maximum light of $\rm{EH~Lib}$ from new ground-based \textit{R}-band observations. $T_{\rm{max}}$ is in BJD-2450000, O-C is in days. Epoch and O−C are derived from the ephemeris formula in Equation~\eqref{eq:equation7}.}
\label{tab:TabelA1}
\begin{tabular}{ccrccrccrccr}
\hline
\textbf{$T_{\rm{max}}$} & Epoch & \multicolumn{1}{c}{O-C} & \textbf{$T_{\rm{max}}$} & Epoch & \multicolumn{1}{c}{O-C} & \textbf{$T_{\rm{max}}$} & Epoch & \multicolumn{1}{c}{O-C} & \textbf{$T_{\rm{max}}$} & Epoch & \multicolumn{1}{c}{O-C} \\
\hline
6703.40821 & 242946 & -0.00005 & 6730.99405 & 243258 & 0.00085 & 7052.46364 & 246894 & -0.00021 & 7437.32783 & 251247 & 0.00104 \\
6704.38111 & 242957 & 0.00030 & 6731.34709 & 243262 & 0.00024 & 7053.43703 & 246905 & 0.00064 & 7440.33358 & 251281 & 0.00073 \\
6705.44244 & 242969 & 0.00067 & 6732.32008 & 243273 & 0.00068 & 7054.40930 & 246916 & 0.00036 & 7441.30600 & 251292 & 0.00061 \\
6723.92075 & 243178 & 0.00061 & 6733.02777 & 243281 & 0.00106 & 7056.44270 & 246939 & 0.00026 & 7441.39419 & 251293 & 0.00038 \\
6724.00923 & 243179 & 0.00067 & 6734.97284 & 243303 & 0.00104 & 7057.41543 & 246950 & 0.00044 & 7795.04718 & 255293 & 0.00031 \\
6725.95418 & 243201 & 0.00053 & 6736.29821 & 243318 & 0.00021 & 7106.30795 & 247503 & 0.00042 & 7798.93725 & 255337 & 0.00019 \\
6726.04259 & 243202 & 0.00053 & 6736.38675 & 243319 & 0.00034 & 7117.80261 & 247633 & 0.00136 & 8198.38830 & 259855 & 0.00010 \\
6729.31379 & 243239 & 0.00044 & 6737.27107 & 243329 & 0.00053 & 7119.83577 & 247656 & 0.00101 & 8199.36076 & 259866 & 0.00001 \\
6729.40208 & 243240 & 0.00032 & 6761.23047 & 243600 & -0.00007 & 7435.29462 & 251224 & 0.00133 &  &  &  \\
6730.37518 & 243251 & 0.00086 & 6761.31932 & 243601 & 0.00037 & 7435.38199 & 251225 & 0.00029 &  &  &  \\
\hline
\end{tabular}
\end{table*}

\begin{table*}
\centering
\renewcommand{\arraystretch}{1.2}
\setlength{\tabcolsep}{4pt}
\caption{Times of maximum light of $\rm{EH~Lib}$ from TESS observations. $T_{\rm{max}}$ is in BJD-2450000, O-C is in days. Epoch and O−C are derived from the ephemeris formula in Equation~\eqref{eq:equation7}.}
\label{tab:TabelA2}
\begin{tabular}{ccrccrccrccr}
\hline
\textbf{$T_{\rm{max}}$} & Epoch & \multicolumn{1}{c}{O-C} & \textbf{$T_{\rm{max}}$} & Epoch & \multicolumn{1}{c}{O-C} & \textbf{$T_{\rm{max}}$} & Epoch & \multicolumn{1}{c}{O-C} & \textbf{$T_{\rm{max}}$} & Epoch & \multicolumn{1}{c}{O-C} \\
\hline
9693.01456 & 276760 & 0.00008  & 9700.35275 & 276843 & -0.00003 & 9703.88934 & 276883 & 0.00003  & 9714.23368 & 277000 & 0.00002  \\
9693.10299 & 276761 & 0.00010  & 9700.44118 & 276844 & -0.00001 & 9703.97777 & 276884 & 0.00005  & 9714.32211 & 277001 & 0.00004  \\
9693.19126 & 276762 & -0.00004 & 9700.52960 & 276845 & 0.00000  & 9704.06604 & 276885 & -0.00009 & 9714.41053 & 277002 & 0.00005  \\
9693.27969 & 276763 & -0.00003 & 9700.61803 & 276846 & 0.00001  & 9704.15447 & 276886 & -0.00008 & 9714.49896 & 277003 & 0.00006  \\
9693.36811 & 276764 & -0.00002 & 9700.70646 & 276847 & 0.00003  & 9704.24305 & 276887 & 0.00009  & 9714.58738 & 277004 & 0.00007  \\
9693.45654 & 276765 & 0.00000  & 9700.79488 & 276848 & 0.00004  & 9704.33132 & 276888 & -0.00006 & 9714.67565 & 277005 & -0.00007 \\
9693.54497 & 276766 & 0.00001  & 9700.88315 & 276849 & -0.00010 & 9704.41974 & 276889 & -0.00004 & 9714.76408 & 277006 & -0.00006 \\
9693.63339 & 276767 & 0.00003  & 9700.97158 & 276850 & -0.00009 & 9704.50832 & 276890 & 0.00012  & 9714.85250 & 277007 & -0.00005 \\
9693.72182 & 276768 & 0.00004  & 9701.06001 & 276851 & -0.00007 & 9704.59675 & 276891 & 0.00014  & 9714.94108 & 277008 & 0.00011  \\
9693.81009 & 276769 & -0.00010 & 9701.14859 & 276852 & 0.00009  & 9704.68518 & 276892 & 0.00015  & 9715.02950 & 277009 & 0.00013  \\
9693.89883 & 276770 & 0.00022  & 9701.23701 & 276853 & 0.00011  & 9705.92282 & 276906 & 0.00001  & 9715.11777 & 277010 & -0.00002 \\
9693.98710 & 276771 & 0.00008  & 9701.32544 & 276854 & 0.00012  & 9706.01125 & 276907 & 0.00002  & 9715.20620 & 277011 & -0.00001 \\
9694.07537 & 276772 & -0.00006 & 9701.41371 & 276855 & -0.00002 & 9706.09967 & 276908 & 0.00003  & 9715.29462 & 277012 & 0.00001  \\
9694.16395 & 276773 & 0.00011  & 9701.50229 & 276856 & 0.00015  & 9706.18810 & 276909 & 0.00005  & 9715.38305 & 277013 & 0.00002  \\
9698.05397 & 276817 & -0.00007 & 9701.59057 & 276857 & 0.00000  & 9706.27652 & 276910 & 0.00006  & 9715.47147 & 277014 & 0.00003  \\
9698.14255 & 276818 & 0.00010  & 9701.67899 & 276858 & 0.00002  & 9706.36479 & 276911 & -0.00009 & 9715.55990 & 277015 & 0.00004  \\
9698.23082 & 276819 & -0.00004 & 9701.76742 & 276859 & 0.00003  & 9706.45337 & 276912 & 0.00008  & 9715.64832 & 277016 & 0.00005  \\
9698.31925 & 276820 & -0.00003 & 9701.85569 & 276860 & -0.00011 & 9706.54180 & 276913 & 0.00009  & 9715.73675 & 277017 & 0.00006  \\
9698.40767 & 276821 & -0.00001 & 9701.94427 & 276861 & 0.00006  & 9706.63022 & 276914 & 0.00010  & 9715.82517 & 277018 & 0.00007  \\
9698.49625 & 276822 & 0.00016  & 9702.03254 & 276862 & -0.00008 & 9706.71849 & 276915 & -0.00004 & 9715.91359 & 277019 & 0.00008  \\
9698.58468 & 276823 & 0.00017  & 9702.12112 & 276863 & 0.00008  & 9706.80707 & 276916 & 0.00013  & 9716.00202 & 277020 & 0.00009  \\
9698.67295 & 276824 & 0.00003  & 9702.20955 & 276864 & 0.00010  & 9706.89534 & 276917 & -0.00002 & 9716.09029 & 277021 & -0.00005 \\
9698.76138 & 276825 & 0.00004  & 9702.29782 & 276865 & -0.00004 & 9706.98377 & 276918 & 0.00000  & 9716.17871 & 277022 & -0.00004 \\
9698.84965 & 276826 & -0.00010 & 9702.38625 & 276866 & -0.00003 & 9707.07235 & 276919 & 0.00016  & 9716.26714 & 277023 & -0.00003 \\
9698.93808 & 276827 & -0.00009 & 9702.47468 & 276867 & -0.00002 & 9707.16062 & 276920 & 0.00002  & 9716.35556 & 277024 & -0.00002 \\
9699.02650 & 276828 & -0.00007 & 9702.56326 & 276868 & 0.00015  & 9707.24920 & 276921 & 0.00019  & 9716.44399 & 277025 & 0.00000  \\
9699.11493 & 276829 & -0.00006 & 9702.65153 & 276869 & 0.00001  & 9712.90747 & 276985 & 0.00001  & 9716.53241 & 277026 & 0.00001  \\
9699.20351 & 276830 & 0.00011  & 9702.73980 & 276870 & -0.00013 & 9712.99589 & 276986 & 0.00002  & 9716.62084 & 277027 & 0.00002  \\
9699.29194 & 276831 & 0.00012  & 9702.82823 & 276871 & -0.00012 & 9713.08416 & 276987 & -0.00012 & 9716.70911 & 277028 & -0.00012 \\
9699.38021 & 276832 & -0.00002 & 9702.91681 & 276872 & 0.00005  & 9713.17274 & 276988 & 0.00004  & 9716.79753 & 277029 & -0.00011 \\
9699.46848 & 276833 & -0.00016 & 9703.00523 & 276873 & 0.00006  & 9713.26117 & 276989 & 0.00005  & 9716.88611 & 277030 & 0.00005  \\
9699.55706 & 276834 & 0.00001  & 9703.09366 & 276874 & 0.00007  & 9713.34959 & 276990 & 0.00006  & 9716.97438 & 277031 & -0.00009 \\
9699.64549 & 276835 & 0.00002  & 9703.18209 & 276875 & 0.00009  & 9713.43786 & 276991 & -0.00008 & 9717.06311 & 277032 & 0.00023  \\
9699.73392 & 276836 & 0.00003  & 9703.27051 & 276876 & 0.00010  & 9713.52613 & 276992 & -0.00022 & 9717.15138 & 277033 & 0.00008  \\
9699.82234 & 276837 & 0.00005  & 9703.35894 & 276877 & 0.00011  & 9713.61471 & 276993 & -0.00006 & 9717.23965 & 277034 & -0.00006 \\
9699.91077 & 276838 & 0.00006  & 9703.44736 & 276878 & 0.00012  & 9713.70329 & 276994 & 0.00011  & 9717.32823 & 277035 & 0.00011  \\
9699.99920 & 276839 & 0.00007  & 9703.53564 & 276879 & -0.00002 & 9713.79156 & 276995 & -0.00003 & 9717.41650 & 277036 & -0.00004 \\
9700.08747 & 276840 & -0.00007 & 9703.62406 & 276880 & 0.00000  & 9713.87998 & 276996 & -0.00002 & 9717.50492 & 277037 & -0.00003 \\
9700.17590 & 276841 & -0.00005 & 9703.71264 & 276881 & 0.00016  & 9713.96825 & 276997 & -0.00016 &            &        &          \\
9700.26432 & 276842 & -0.00004 & 9703.80092 & 276882 & 0.00002  & 9714.14526 & 276999 & 0.00001  &            &        &          \\
\hline
\end{tabular}
\end{table*}

\clearpage
\begin{table*}
\centering
\renewcommand{\arraystretch}{1.1}
\setlength{\tabcolsep}{20pt}
\caption{Times of maximum light of $\rm{EH~Lib}$ reported in previous literature. $T_{\rm{max}}$ is in BJD-2400000, O-C is in days. Epoch and O−C are derived from the ephemeris formula in Equation~\eqref{eq:equation7}. Det: detector (pg=photograph, pe=photoelectric photometer).}
\label{tab:TabelA3}
\begin{tabular}{crrcl}
\hline
\textbf{$T_{\rm{max}}$} & \multicolumn{1}{c}{Epoch} & \multicolumn{1}{c}{O-C} & Det. & \multicolumn{1}{c}{Reference} \\
\hline
33438.60796 & -20191 & 0.00154  & pe  & \cite{1950PASP...62..166C} \\
33711.71687 & -17102 & 0.00186  & pg  & \cite{1952AJ.....57Q..64A} \\
33737.70937 & -16808 & 0.00086  & pg  & \cite{1952AJ.....57Q..64A} \\
33743.72237 & -16740 & 0.00176  & pg  & \cite{1952AJ.....57Q..64A} \\
35223.75995 & 0      & 0.00125  & pe  & \cite{1957AJ.....62..108F} \\
35223.84815 & 1      & 0.00104  & pe  & \cite{1957AJ.....62..108F} \\
35223.93755 & 2      & 0.00203  & pe  & \cite{1957AJ.....62..108F} \\
35225.79355 & 23     & 0.00135  & pe  & \cite{1957AJ.....62..108F} \\
35243.74175 & 226    & 0.00165  & pe  & \cite{1957AJ.....62..108F} \\
35243.83015 & 227    & 0.00164  & pe  & \cite{1957AJ.....62..108F} \\
35622.68024 & 4512   & 0.00088  & pe  & \cite{1957AJ.....62..108F} \\
35622.76894 & 4513   & 0.00116  & pe  & \cite{1957AJ.....62..108F} \\
36996.44464 & 20050  & -0.00006 & pe  & \cite{1961Obs....81..199S} \\
37054.26775 & 20704  & 0.00077  & pe  & \cite{1961Obs....81..199S} \\
37054.35555 & 20705  & 0.00015  & pe  & \cite{1961Obs....81..199S} \\
37075.22255 & 20941  & 0.00162  & pe  & \cite{1961Obs....81..199S} \\
37075.31125 & 20942  & 0.00191  & pe  & \cite{1961Obs....81..199S} \\
37075.39855 & 20943  & 0.00080  & pe  & \cite{1961Obs....81..199S} \\
37077.16735 & 20963  & 0.00133  & pe  & \cite{1961Obs....81..199S} \\
37082.20575 & 21020  & 0.00018  & pe  & \cite{1961Obs....81..199S} \\
37105.19305 & 21280  & 0.00003  & pe  & \cite{1961Obs....81..199S} \\
37403.32207 & 24652  & -0.00049 & pe  & \cite{1961Obs....81..199S} \\
37403.41187 & 24653  & 0.00090  & pe  & \cite{1961Obs....81..199S} \\
37408.36277 & 24709  & 0.00065  & pe  & \cite{1961Obs....81..199S} \\
37410.30777 & 24731  & 0.00056  & pe  & \cite{1961Obs....81..199S} \\
37412.34057 & 24754  & -0.00014 & pe  & \cite{1961Obs....81..199S} \\
38441.03042 & 36389  & 0.00135  & pe  & \cite{1966CoLPL...5....3F} \\
38467.02292 & 36683  & 0.00035  & pe  & \cite{1966CoLPL...5....3F} \\
37114.30135 & 21383  & 0.00176  & pe  & \cite{1966BAN....18..387O} \\
37116.24535 & 21405  & 0.00067  & pe  & \cite{1966BAN....18..387O} \\
37116.33435 & 21406  & 0.00126  & pe  & \cite{1966BAN....18..387O} \\
42162.51863 & 78481  & -0.00168 & pe  & \cite{1977IBVS.1310....1K} \\
42182.41383 & 78706  & 0.00054  & pe  & \cite{1977IBVS.1310....1K} \\
42541.45935 & 82767  & -0.00022 & pe  & \cite{1977IBVS.1310....1K} \\
42544.37635 & 82800  & -0.00086 & pe  & \cite{1977IBVS.1310....1K} \\
42544.46585 & 82801  & 0.00023  & pe  & \cite{1977IBVS.1310....1K} \\
42577.44425 & 83174  & 0.00048  & pe  & \cite{1977IBVS.1310....1K} \\
42871.50746 & 86500  & 0.00117  & pe  & \cite{1977IBVS.1310....1K} \\
42871.59386 & 86501  & -0.00085 & pe  & \cite{1977IBVS.1310....1K} \\
42872.47786 & 86511  & -0.00098 & pe  & \cite{1977IBVS.1310....1K} \\
42872.56636 & 86512  & -0.00089 & pe  & \cite{1977IBVS.1310....1K} \\
42874.51126 & 86534  & -0.00108 & pe  & \cite{1977IBVS.1310....1K} \\
43249.64837 & 90777  & -0.00147 & pe  & \cite{1979AAS...36...51G} \\
43254.60017 & 90833  & -0.00081 & pe  & \cite{1979AAS...36...51G} \\
43255.57337 & 90844  & -0.00016 & pe  & \cite{1979AAS...36...51G} \\
43255.66187 & 90845  & -0.00007 & pe  & \cite{1979AAS...36...51G} \\
43256.63347 & 90856  & -0.00102 & pe  & \cite{1979AAS...36...51G} \\
\hline
\end{tabular}
\end{table*}

\begin{table*}
\ContinuedFloat
\centering
\renewcommand{\arraystretch}{1.1}
\setlength{\tabcolsep}{20pt}
\begin{tabular}{crrcl}
\hline
\textbf{$T_{\rm{max}}$} & \multicolumn{1}{c}{Epoch} & \multicolumn{1}{c}{O-C} & Det. & \multicolumn{1}{c}{Reference} \\
\hline
43274.58127 & 91059  & -0.00111 & pe  & \cite{1979AAS...36...51G} \\
43287.57837 & 91206  & -0.00076 & pe  & \cite{1979AAS...36...51G} \\
41476.43317 & 70721  & -0.00018 & pe  & \cite{1980CoKon..74....1M} \\
42159.51343 & 78447  & -0.00083 & pe  & \cite{1980CoKon..74....1M} \\
42548.44325 & 82846  & -0.00097 & pe  & \cite{1980CoKon..74....1M} \\
42871.50566 & 86500  & -0.00063 & pe  & \cite{1980CoKon..74....1M} \\
43957.57376 & 98784  & -0.00110 & pe  & \cite{1980CoKon..74....1M} \\
44378.15586 & 103541 & -0.00092 & pe  & \cite{1981AcASn..22..279J} \\
44379.12926 & 103552 & -0.00007 & pe  & \cite{1981AcASn..22..279J} \\
44379.21766 & 103553 & -0.00008 & pe  & \cite{1981AcASn..22..279J} \\
44406.09416 & 103857 & -0.00122 & pe  & \cite{1981AcASn..22..279J} \\
44406.18316 & 103858 & -0.00063 & pe  & \cite{1981AcASn..22..279J} \\
44620.40855 & 106281 & -0.00058 & pe  & \cite{1981AcASn..22..279J} \\
46146.50910 & 123542 & -0.00144 & pe  & \cite{1985IBVS.2810....1H} \\
46146.59730 & 123543 & -0.00165 & pe  & \cite{1985IBVS.2810....1H} \\
46147.48190 & 123553 & -0.00118 & pe  & \cite{1985IBVS.2810....1H} \\
46147.57010 & 123554 & -0.00139 & pe  & \cite{1985IBVS.2810....1H} \\
46148.45410 & 123564 & -0.00153 & pe  & \cite{1985IBVS.2810....1H} \\
46148.54250 & 123565 & -0.00154 & pe  & \cite{1985IBVS.2810....1H} \\
46151.54870 & 123599 & -0.00139 & pe  & \cite{1985IBVS.2810....1H} \\
45433.98716 & 115483 & -0.00085 & pe  & \cite{1986PASP...98..651J} \\
45439.91076 & 115550 & -0.00094 & pe  & \cite{1986PASP...98..651J} \\
45440.97156 & 115562 & -0.00110 & pe  & \cite{1986PASP...98..651J} \\
45441.94436 & 115573 & -0.00084 & pe  & \cite{1986PASP...98..651J} \\
45458.91976 & 115765 & -0.00079 & pe  & \cite{1986PASP...98..651J} \\
45477.75166 & 115978 & -0.00092 & pe  & \cite{1986PASP...98..651J} \\
45477.84036 & 115979 & -0.00063 & pe  & \cite{1986PASP...98..651J} \\
45519.74828 & 116453 & -0.00061 & pe  & \cite{1986PASP...98..651J} \\
45520.72008 & 116464 & -0.00135 & pe  & \cite{1986PASP...98..651J} \\
46094.34669 & 122952 & -0.00001 & pe  & \cite{1992IBVS.3769....1Y} \\
46095.31889 & 122963 & -0.00036 & pe  & \cite{1992IBVS.3769....1Y} \\
46095.40759 & 122964 & -0.00007 & pe  & \cite{1992IBVS.3769....1Y} \\
46179.04430 & 123910 & -0.00232 & pe  & \cite{1992IBVS.3769....1Y} \\
46179.13370 & 123911 & -0.00133 & pe  & \cite{1992IBVS.3769....1Y} \\
46179.22200 & 123912 & -0.00144 & pe  & \cite{1992IBVS.3769....1Y} \\
46211.13890 & 124273 & -0.00173 & pe  & \cite{1992IBVS.3769....1Y} \\
46211.22730 & 124274 & -0.00174 & pe  & \cite{1992IBVS.3769....1Y} \\
56753.89238 & 243517 & 0.00014  & CCD & \cite{2017IBVS.6231....1P} \\
56753.98078 & 243518 & 0.00013  & CCD & \cite{2017IBVS.6231....1P} \\
57088.80307 & 247305 & 0.00137  & CCD & \cite{2017IBVS.6231....1P} \\
57114.79547 & 247599 & 0.00027  & CCD & \cite{2017IBVS.6231....1P} \\
57172.79276 & 248255 & -0.00154 & CCD & \cite{2017IBVS.6231....1P} \\
57175.79976 & 248289 & -0.00059 & CCD & \cite{2017IBVS.6231....1P} \\
57177.83176 & 248312 & -0.00209 & CCD & \cite{2017IBVS.6231....1P} \\
57459.87287 & 251502 & 0.00069  & CCD & \cite{2017IBVS.6231....1P} \\
57460.84487 & 251513 & 0.00014  & CCD & \cite{2017IBVS.6231....1P} \\
\hline
\end{tabular}
\end{table*}
    
\clearpage
\bibliography{EHLib}{}
\bibliographystyle{aasjournalv7.1}



\end{document}